\documentclass[journal=jacsat,manuscript=article]{achemso}
\setkeys{acs}{articletitle=true}

\usepackage[version=3]{mhchem} % Formula subscripts using \ce{}
\usepackage[T1]{fontenc}       % Use modern font encodings
\usepackage{listings}
\usepackage{multirow}
\usepackage{lineno}
\usepackage{hyperref}
\usepackage{amsmath}
\usepackage{amsthm}
\usepackage{amssymb}
\usepackage{xcolor}
\usepackage[colorinlistoftodos]{todonotes}
\usepackage{pifont}
\usepackage{algorithmicx}
\usepackage{algcompatible}
\usepackage{algorithm}
\usepackage{url}
\usepackage{array,booktabs}
\definecolor{bg}{rgb}{0.95,0.95,0.95}
\newcommand{\beginsupplement}{%
        \setcounter{table}{0}
        \renewcommand{\thetable}{S\arabic{table}}%
        \setcounter{figure}{0}
        \renewcommand{\thefigure}{S\arabic{figure}}%
     }

\author{Tijana Radivojevi\'c}
\affiliation{DOE Agile BioFoundry, Emeryville, CA, USA.}
\alsoaffiliation{Biological Systems and Engineering Division, Lawrence Berkeley National Laboratory, Berkeley, CA, USA.}

\author{Zak Costello}
\affiliation{Biofuels and Bioproducts Division, DOE Joint BioEnergy Institute, Emeryville, CA, USA.}
\alsoaffiliation{DOE Agile BioFoundry, Emeryville, CA, USA.}
\alsoaffiliation{Biological Systems and Engineering Division, Lawrence Berkeley National Laboratory, Berkeley, CA, USA.}
\author{Kenneth Workman}
\affiliation{DOE Agile BioFoundry, Emeryville, CA, USA.}
\alsoaffiliation{Biological Systems and Engineering Division, Lawrence Berkeley National Laboratory, Berkeley, CA, USA.}
\alsoaffiliation{Department of Bioengineering, University of California, Berkeley, CA, USA}

\author{Hector Garcia Martin}
\email{hgmartin@lbl.gov}
\affiliation{Biofuels and Bioproducts Division, DOE Joint BioEnergy Institute, Emeryville, CA, USA.}
\alsoaffiliation{DOE Agile BioFoundry, Emeryville, CA, USA.}
\alsoaffiliation{Biological Systems and Engineering Division, Lawrence Berkeley National Laboratory, Berkeley, CA, USA.}
\alsoaffiliation{BCAM, Basque Center for Applied Mathematics, Bilbao, Spain.}

\title[ART]
  {ART: A machine learning Automated Recommendation Tool for synthetic biology}


\begin{document}

%%%%%%%%%%%%%%%%%%%%%%%%%%%%%%%%%%%%%%%%%%%%%%%%%%%%%%%%%%%%%%%%%%%%%
%% The "tocentry" environment can be used to create an entry for the
%% graphical table of contents. It is given here as some journals
%% require that it is printed as part of the abstract page. It will
%% be automatically moved as appropriate.
%%%%%%%%%%%%%%%%%%%%%%%%%%%%%%%%%%%%%%%%%%%%%%%%%%%%%%%%%%%%%%%%%%%%%

%\begin{tocentry}

% Some journals require a graphical entry for the Table of Contents.
% This should be laid out ``print ready'' so that the sizing of the
% text is correct.

% Inside the \texttt{tocentry} environment, the font used is Helvetica
% 8\,pt, as required by \emph{Journal of the American Chemical
% Society}.

% The surrounding frame is 9\,cm by 3.5\,cm, which is the maximum
% permitted for  \emph{Journal of the American Chemical Society}
% graphical table of content entries. The box will not resize if the
% content is too big: instead it will overflow the edge of the box.

% This box and the associated title will always be printed on a
% separate page at the end of the document.

% \includegraphics[width=0.8\columnwidth]{ART_abstract.png}

%\end{tocentry}

% \listoftodos

% \todo[size=\small,noline]{Results}
% \todo[size=\small,noline]{Notebooks (Hector and Zak to review at github.com)}
% \todo[size=\small,noline]{Supporting Information}

%%%%%%%%%%%%%%%%%%%%%%%%%%%%%%%%%%%%%%%%%%%%%%%%%%%%%%%%%%%%%%%%%%%%%
%% The abstract environment will automatically gobble the contents
%% if an abstract is not used by the target journal.
%%%%%%%%%%%%%%%%%%%%%%%%%%%%%%%%%%%%%%%%%%%%%%%%%%%%%%%%%%%%%%%%%%%%%

\clearpage

\begin{abstract} 
  Synthetic biology allows us to bioengineer cells to synthesize novel valuable molecules such as renewable biofuels or anticancer drugs. However, traditional synthetic biology approaches involve ad-hoc engineering practices, which lead to long development times. Here, we present the Automated Recommendation Tool (\textsf{ART}), a tool that leverages machine learning and probabilistic modeling techniques to guide synthetic biology in a systematic fashion, without the need for a full mechanistic understanding of the biological system. Using sampling-based optimization, \textsf{ART} provides a set of recommended strains to be built in the next engineering cycle, alongside probabilistic predictions of their production levels. We demonstrate the capabilities of \textsf{ART} on simulated data sets, as well as experimental data from real metabolic engineering projects producing renewable biofuels, hoppy flavored beer without hops, and fatty acids. Finally, we discuss the limitations of this approach, and the practical consequences of the underlying assumptions failing.  
\end{abstract}

\clearpage

%%%%%%%%%%%%%%%%%%%%%%%%%%%%%%%%%%%%%%%%%%%%%%%%%%%%%%%%%%%%%%%%%%%%%
%% Start the main part of the manuscript here.
%%%%%%%%%%%%%%%%%%%%%%%%%%%%%%%%%%%%%%%%%%%%%%%%%%%%%%%%%%%%%%%%%%%%%
\section{Introduction}
Metabolic engineering\cite{stephanopoulos1999metabolic} enables us to bioengineer cells to synthesize novel valuable molecules such as renewable biofuels\cite{beller2015natural, chubukov2016synthetic} or anticancer drugs\cite{ajikumar2010isoprenoid}. The prospects of metabolic engineering to have a positive impact in society are on the rise, as it was considered one of the ``Top Ten Emerging Technologies'' by the World Economic Forum in 2016\cite{cann2016these}. Furthermore, an incoming industrialized biology is expected to improve most human activities: from creating renewable bioproducts and materials, to improving crops and enabling new biomedical applications~\cite{national2015industrialization}. 

However, the practice of metabolic engineering has been far from systematic, which has significantly hindered its overall impact\cite{yadav2012future}. Metabolic engineering has remained a collection of useful demonstrations rather than a systematic practice based on generalizable methods. This limitation has resulted in very long development times: for example, it took 150 person-years of effort to produce the antimalarial precursor artemisinin by Amyris; and 575 person-years of effort for Dupont to generate propanediol\cite{hodgman2012cell}, which is the base for their commercially available Sorona fabric\cite{kurian2005new}. 

Synthetic biology\cite{cameron2014brief} aims to improve genetic and metabolic engineering by applying systematic engineering principles to achieve a previously specified goal. Synthetic biology encompasses, and goes beyond, metabolic engineering: it also involves non-metabolic tasks such as gene drives able to estinguish malaria-bearing mosquitoes\cite{kyrou2018crispr} or engineering microbiomes to replace fertilizers\cite{temme2019methods}. This discipline is enjoying an exponential growth, as it heavily benefits from the byproducts of the genomic revolution: high-throughput multi-omics phenotyping\cite{chen2019automated,fuhrer2015high}, accelerating DNA sequencing\cite{stephens2015big} and synthesis capabilities\cite{ma2012dna}, and CRISPR-enabled genetic editing\cite{doudna2014new}. This exponential growth is reflected in the private investment in the field, which has totalled $\sim$\$12B in the 2009-2018 period and is rapidly accelerating ($\sim$\$2B in 2017 to $\sim$\$4B in 2018)~\cite{cumbers2019}.   

One of the synthetic biology engineering principles used to improve metabolic engineering is the Design-Build-Test-Learn (DBTL~\cite{petzold2015analytics, nielsen2016engineering}) cycle---a loop used recursively to obtain a design that satisfies the desired specifications (e.g.\ a particular titer, rate, yield or product). The DBTL cycle's first step is to design (D) a biological system expected to meet the desired outcome. That design is built (B) in the next phase from DNA parts into an appropriate microbial chassis using synthetic biology tools. The next phase involves testing (T) whether the built biological system indeed works as desired in the original design, via a variety of assays: e.g. measurement of production or/and `omics (transcriptomics, proteomics, metabolomics) data profiling. It is extremely rare that the first design behaves as desired, and further attempts are typically needed to meet the desired specification. The Learn (L) step leverages the data previously generated to inform the next Design step so as to converge to the desired specification faster than through a random search process. 

The Learn phase of the DBTL cycle has traditionally been the most weakly supported and developed~\cite{nielsen2016engineering}, despite its critical importance to accelerate the full cycle. The reasons are multiple, although their relative importance is not entirely clear. Arguably, the main drivers of the lack of emphasis on the L phase are: the lack of predictive power for biological systems behavior\cite{gardner2013synthetic}, the reproducibility problems plaguing biological experiments\cite{chubukov2016synthetic,prinz2011believe,baker20161,begley2012drug}, and the traditionally moderate emphasis on mathematical training for synthetic biologists.

Machine learning (ML) arises as an effective tool to predict biological system behavior and empower the Learn phase, enabled by emerging high-throughput phenotyping technologies \cite{Carbonell2019}. Machine learning has been used to produce driverless cars\cite{thrun2010toward}, automate language translation\cite{wu2016google}, predict %sexual orientation
sensitive personal attributes from Facebook profiles\cite{kosinski2013private}, predict pathway dynamics\cite{costello2018machine}, optimize pathways through translational control\cite{jervis2018machine}, diagnose skin cancer\cite{esteva2017dermatologist}, detect tumors in breast tissues\cite{paeng2017unified}, predict DNA and RNA protein-binding sequences\cite{alipanahi2015predicting}, drug side effects\cite{shaked2016metabolic} and antibiotic mechanisms of action\cite{yang2019white}. However, the practice of machine learning requires statistical and mathematical expertise that is scarce and highly competed for in other fields~\cite{metz2018ai}.% {\color{blue}(throughout industry and academia)}.

{\color{black}In this paper}, we provide a tool that leverages machine learning for synthetic biology's purposes: the Automated Recommendation Tool (\textsf{ART}). \textsf{ART} combines the widely-used and general-purpose open source scikit-learn library~\cite{pedregosa2011scikit} with a novel Bayesian\cite{Gelman2003} ensemble approach, in a manner that adapts to the particular needs of synthetic biology projects: e.g.\ low number of training instances, recursive DBTL cycles, and the need for uncertainty quantification. The data sets collected in the synthetic biology field are typically not large enough to allow for the use of deep learning ($<100$ instances), but our ensemble model will be able to integrate this approach when high-throughput data generation\cite{batth2014targeted, fuhrer2015high} and automated data collection\cite{heinemann2017chip} become widely used in the future. \textsf{ART} provides machine learning capabilities in an easy-to-use and intuitive manner, and is able to guide synthetic biology efforts in an effective way. 

We showcase the efficacy of \textsf{ART} in guiding synthetic biology by mapping --omics data to production through four different examples: {\color{black}one} test case with simulated data and three real cases of metabolic engineering. In all these cases we assume that the -omics data (proteomics in these examples, but it could be any other type: transcriptomics, metabolomics, etc.) can be predictive of the final production (response), and that we have enough control over the system so as to produce any new recommended input. The test case permits us to explore how the algorithm performs when applied to systems that present different levels of difficulty when being ``learnt'', as well as the effectiveness of using several DTBL cycles. The real metabolic engineering cases involve data sets from published metabolic engineering projects: renewable biofuel production, yeast bioengineering to recreate the flavor of hops in beer, and fatty alcohols synthesis. These projects illustrate what to expect under different typical metabolic engineering situations: high/low coupling of the heterologous pathway to host metabolism, complex/simple pathways, high/low number of conditions, high/low difficulty in learning pathway behavior. We find that  \textsf{ART}'s ensemble approach can successfully guide the bioengineering process even in the absence of quantitatively accurate predictions. Furthermore, \textsf{ART}'s ability to quantify uncertainty is crucial to gauge the reliability of predictions and effectively guide recommendations towards the least known part of the phase space. These experimental metabolic engineering cases also illustrate how applicable the underlying assumptions are, and what happens when they fail.

In sum, \textsf{ART} provides a tool specifically tailored to the synthetic biologist's needs in order to leverage the power of machine learning to enable predictable biology. This combination of synthetic biology with machine learning and automation has the potential to revolutionize bioengineering~\cite{Carbonell2019,hamedirad2019towards,hase2019next} by enabling effective inverse design. This paper is written so as to be accessible to both the machine learning and synthetic biology readership, with the intention of providing a much needed bridge between these two very different collectives. Hence, we apologize if we put emphasis on explaining basic machine learning or synthetic biology concepts---they will surely be of use to a part of the readership.

\section{Methods}

\subsection{Key capabilities}\label{Sec:Key_cap}

\begin{figure}
 \begin{center}
 \includegraphics[width=0.8\columnwidth]{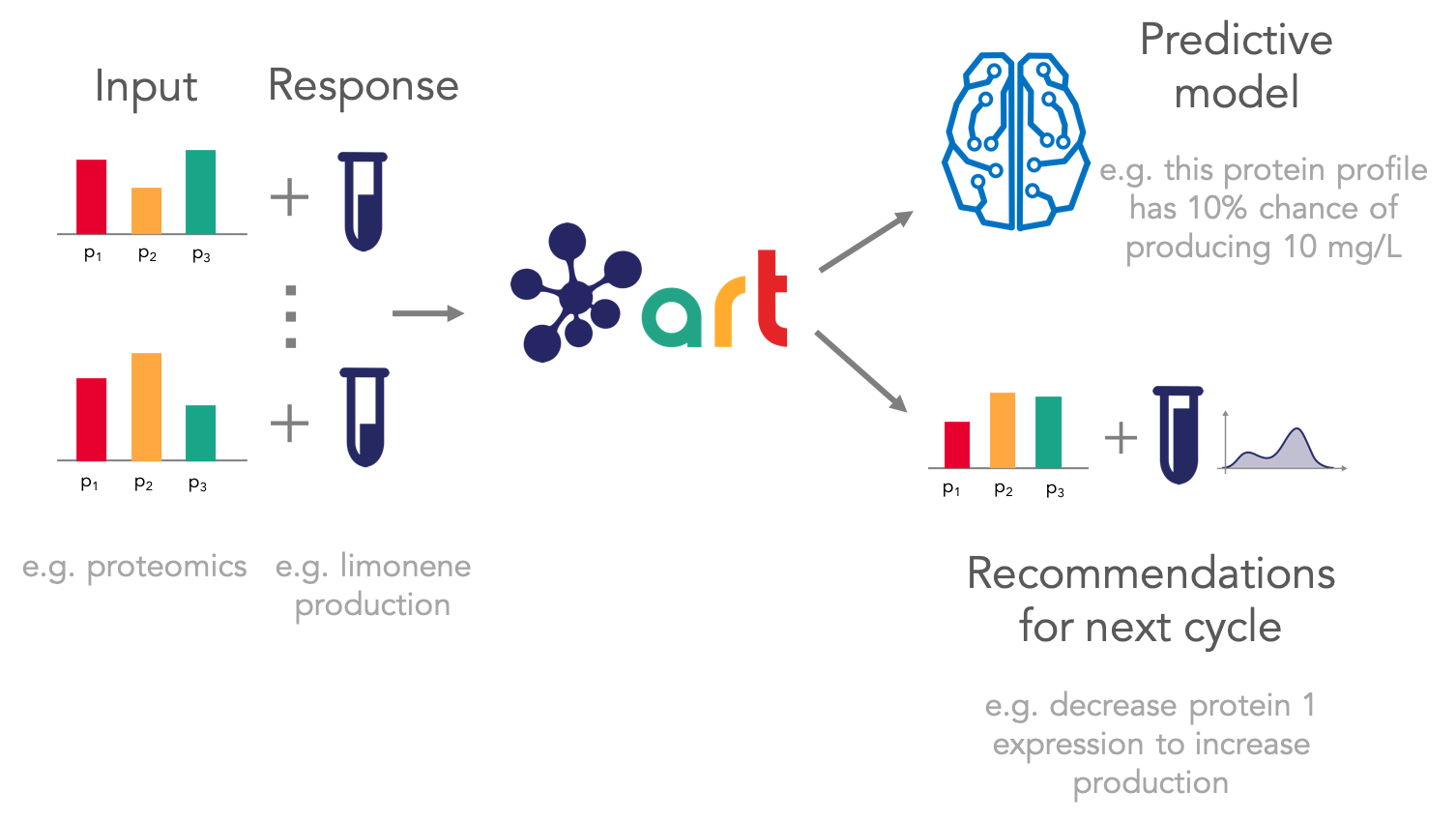}
 \end{center}
 \caption{\textbf{\textsf{ART} predicts the response from the input and provides recommendations for the next cycle.} \textsf{ART} uses experimental data to i) build a probabilistic predictive model that predicts response (e.g. production) from input variables (e.g. proteomics), and ii) uses this model to provide a set of recommended designs for the next experiment, along with the probabilistic predictions of the response.}
  \label{Fig:overview}
\end{figure}

\textsf{ART} leverages machine learning to improve the efficacy of bioengineering microbial strains for the production of desired bioproducts (Fig.~\ref{Fig:overview}). \textsf{ART} gets trained on available data to produce a model capable of predicting the response variable (e.g. production of the jet fuel limonene) from the input data (e.g. proteomics data, or any other type of data that can be expressed as a vector). Furthermore, \textsf{ART} uses this model to recommend new inputs (e.g. proteomics profiles) that are predicted to reach our desired goal (e.g. improve production). As such, \textsf{ART} bridges the Learn and Design phases of a DBTL cycle. 

\textsf{ART} can import data directly from Experimental Data Depot \cite{Morrell2017}, an online tool where experimental data and metadata are stored in a standardized manner. Alternatively, \textsf{ART} can import EDD-style .csv files, which use the nomenclature and structure of EDD exported files.

By training on the provided data set, \textsf{ART} builds a predictive model for the response as a function of the input variables. Rather than predicting point estimates of the output variable, \textsf{ART} provides the full probability distribution of the predictions. This rigorous quantification of uncertainty enables a principled way to test hypothetical scenarios in-silico, and to guide design of experiments in the next DBTL cycle. The Bayesian framework chosen to provide the uncertainty quantification is particularly tailored to the type of problems most often encountered in metabolic engineering: sparse data which is expensive and time consuming to generate. 

With a predictive model at hand, \textsf{ART} can provide a set of recommendations expected to produce a desired outcome, as well as probabilistic predictions of the associated response. \textsf{ART} supports the following typical metabolic engineering objectives: maximization of the production of a target molecule (e.g.\ to increase Titer, Rate and Yield, TRY), its minimization (e.g.\ to decrease the toxicity), as well as specification objectives (e.g.\ to reach specific level of a target molecule for a desired beer taste profile). Furthermore, \textsf{ART} leverages the probabilistic model to estimate the probability that at least one of the provided recommendations is successful (e.g.\ it improves the best production obtained so far), and derives how many strain constructions would be required for a reasonable chance to achieve the desired goal.

While \textsf{ART} can be applied to problems with multiple output variables of interest, it currently supports only the same type of objective for all output variables. Hence, it does not {\color{black}yet} support maximization of one target molecule along with minimization of another (see "Success probability calculation" in the supplementary material).

% \clearpage

\subsection{Mathematical methodology}

\subsubsection{Learning from data: a predictive model through machine learning and a novel Bayesian ensemble approach}

\begin{figure}[h!]
\begin{center}
\includegraphics[width=1.\columnwidth]{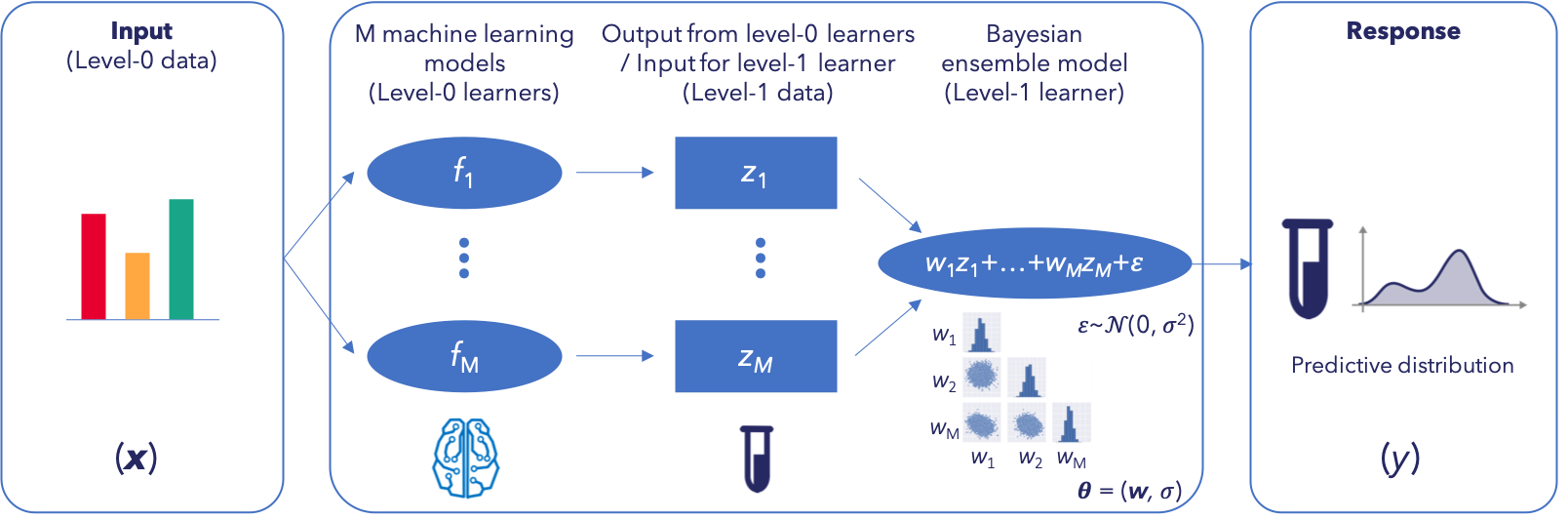}
\end{center}
\caption{\textbf{\textsf{ART} provides a probabilistic predictive model of the response (e.g.\ production).} \textsf{ART} combines several machine learning models from the scikit-learn library with a novel Bayesian approach to predict the probability distribution of the output. The input to \textsf{ART} is proteomics data (or any other input data in vector format: transcriptomics, gene copy, etc.), which we call level-0 data. This level-0 data is used as input for a variety of machine learning models from the scikit-learn library (level-0 learners) that produce a prediction of production for each model ($z_i$). These predictions (level-1 data) are used as input for the Bayesian ensemble model (level-1 learner), which weights these predictions differently depending on its ability to predict the training data. The weights $w_i$ and the variance $\sigma$ are characterized through probability distributions, giving rise to a final prediction in the form of a full probability distribution of response levels.}
\label{Fig:Predictive_model}
\end{figure}

By learning the underlying regularities in experimental data, machine learning can provide predictions without a detailed mechanistic understanding (Fig.~\ref{Fig:Predictive_model}). Training data is used to statistically link an input (i.e.\ features or independent variables) to an output (i.e.\ response or dependent variables) through models that are expressive enough to represent almost any relationship. After this training, the models can be used to predict the outputs for inputs that the model has never seen before.

Model selection is a significant challenge in machine learning, since there is a large variety of models available for learning the relationship between response and input, but none of them is optimal for all learning tasks \cite{Wolpert1996}. Furthermore, each model features hyperparameters (i.e.\ parameters that are set before the training process) that crucially affect the quality of the predictions (e.g.\ number of trees for random forest or degree of polynomials in polynomial regression), and finding their optimal values is not trivial.

We have sidestepped the challenge of model selection by using an ensemble model approach. This approach takes the input of various different models and has them ``vote'' for a particular prediction. Each of the ensemble members is trained to perform the same task and their predictions are combined to achieve an improved performance. The examples of the random forest \cite{Ho1995} or the super learner algorithm \cite{vanderLaan2007} have shown that simple models can be significantly improved by using a set of them (e.g.\ several types of decision trees in a random forest algorithm). Ensemble model typically either use a set of different models (heterogeneous case) or the same models with different parameters (homogeneous case). We have chosen a heterogeneous ensemble learning approach that uses reasonable hyperparameters for each of the model types, rather than specifically tuning hyperparameters for each of them. 

\textsf{ART} uses a novel probabilistic ensemble approach where the weight of each ensemble model is considered a random variable, with a probability distribution inferred by the available data. Unlike other approaches \cite{Hoeting1999, Monteith2011, Yao2018, Chipman2006}, this method does not require the individual models to be probabilistic in nature, hence allowing us to fully exploit the popular scikit-learn library to increase accuracy by leveraging a diverse set of models (see ``Related work and novelty of our ensemble approach'' in the supplementary material). Our weighted ensemble model approach produces a simple, yet powerful, way to quantify both epistemic and aleatoric uncertainty---a critical capability when dealing with small data sets and a crucial component of AI in biological research\cite{begoli2019need}. Here we describe our approach for the single response variable problems, whereas the multiple variables case can be found in the ``Multiple response variables'' section in the supplementary material. Using a common notation in ensemble modeling we define the following levels of data and learners (see Fig.~\ref{Fig:Predictive_model}):

\begin{itemize}
\item \textit{Level-0 data} ($\mathcal D$) represent the historical data consisting of $N$ known instances of inputs and responses, i.e.\ $\mathcal D=\{(\mathbf x_n,y_n),n=1,\dots,N\}$, where $\mathbf x \in \mathcal X \subseteq \mathbb R^D$ is the input comprised of $D$ features and $y \in \mathbb R$ is the associated response variable. For the sake of cross-validation, the \textit{level-0 data} are further divided into validation ($\mathcal D^{(k)}$) and training sets ($\mathcal D^{(-k)}$). $\mathcal D^{(k)} \subset \mathcal D$ is the $k$th fold of a $K$-fold cross-validation obtained by randomly splitting the set $\mathcal D$ into $K$ almost equal parts, and $\mathcal D^{(-k)}=\mathcal D\setminus \mathcal D^{(k)}$ is the set $\mathcal D$ without the $k$th fold $\mathcal D^{(k)}$. Note that these sets do not overlap and cover the full available data; i.e. $\mathcal D^{(k_i)}\cap\mathcal D^{(k_j)}=\emptyset, i\neq j$ and $\cup_i\mathcal D^{(k_i)}=\mathcal D$. 

\item \textit{Level-0 learners} (${f_m}$) consist of $M$ base learning algorithms ${f_m},m=1,\dots,M$ used to learn from level-0 training data $\mathcal D^{(-k)}$. For \textsf{ART}, we have chosen the following eight algorithms from the scikit-learn library: Random Forest, Neural Network, Support Vector Regressor, Kernel Ridge Regressor, K-NN Regressor, Gaussian Process Regressor, Gradient Boosting Regressor, as well as TPOT (tree-based pipeline optimization tool\cite{Olson2016EvoBio}). TPOT uses genetic algorithms to find the combination of the 11 different regressors and 18 different preprocessing algorithms from scikit-learn that, properly tuned, provides the best {\color{black}achieved} cross-validated performance on the training set. 

\item \textit{Level-1 data} ($\mathfrak D_{CV}$) are data derived from $\mathcal D$ by leveraging cross-validated predictions of the level-0 learners. More specifically, level-1 data are given by the set $\mathfrak D_{CV}=\{(\mathbf z_n,y_n),n=1,\dots,N \}$, where $\mathbf z_n=(z_{1n}\dots,z_{Mn})$ are predictions for level-0 data ($\mathbf x_n \in\mathcal D^{(k)}$) of level-0 learners ($f_m^{(-k)}$) trained on observations which are not in fold $k$ ($\mathcal D^{(-k)}$), i.e.\ $z_{mn}=f_m^{(-k)}(\mathbf x_n),m=1,\dots,M$.

%$\dots,z_{Mn})$, $z_{mn}=f_m^{(-k)}(\mathbf x_n),m=1,\dots,M$ are predictions for data $\mathbf x_n \in\mathcal D^{(k)}$ of level-0 learners $f_m^{(-k)}$ trained on observations $\mathcal D^{(-k)}$, which are not in fold $k$.

\item The \textit{level-1 learner} ($F$), or metalearner, is a linear weighted combination of level-0 learners, with weights $w_m$, $m=1,\dots, M$ being random variables that are non-negative and normalized to one. Each $w_m$ can be interpreted as the relative confidence in model $m$. More specifically, given an input $\mathbf x$ the response variable $y$ is modeled as:
\begin{equation}\label{Eq:response_var}
F: \:\:\: y= \boldsymbol{w}^T \mathbf f(\mathbf x)+\varepsilon, \quad \varepsilon \sim \mathcal N(0,\sigma^2),
\end{equation}
where $\boldsymbol{w} = [w_1 \dots w_{M}]^T$ is the vector of weights such that $\sum w_m = 1, w_m\geq 0$, ${\mathbf f}(\mathbf x) = [f_1(\mathbf x) \dots {f_M}(\mathbf x)]^T$ is the vector of level-0 learners, and $\varepsilon$ is a normally distributed error variable with a zero mean and standard deviation $\sigma$. The constraint $\sum w_m=1$ (i.e.\ that the ensemble is a convex combination of the base learners) is empirically motivated but also supported by theoretical considerations~\cite{Breiman1996}.
We denote the unknown ensemble model parameters as $\boldsymbol\theta \equiv(\boldsymbol{w},\sigma)$, constituted of the vector of weights and the Gaussian error standard deviation. The parameters $\boldsymbol\theta$ are obtained by training $F$ on the level-1 data $\mathfrak D_{CV}$ \emph{only}. However, the final model $F$ to be used for generating predictions for new inputs uses  these $\boldsymbol\theta$, inferred from level-1 data $\mathfrak D_{CV}$, and the base learners  $f_m, m=1,\dots,M$  trained on the full original data set $\mathcal D$, rather than only on the level-0 data partitions $\mathcal D^{(-k)}$. This follows the usual procedure in developing ensemble learners \cite{LeDell2015,Aldave2015} in the context of stacking \cite{Breiman1996}. 
\end{itemize}

Rather than providing a single point estimate of ensemble model parameters $\boldsymbol\theta$ that best fit the training data, a Bayesian model provides a joint probability distribution $p(\boldsymbol\theta|\mathcal D)$ which quantifies the probability that a given set of parameters explains the training data. This Bayesian approach makes it possible to not only make predictions for new inputs but also examine the uncertainty in the model. Model parameters $\boldsymbol\theta$ are characterized by full posterior distribution $p(\boldsymbol\theta|\mathcal D)$ that is inferred from level-1 data. Since this distribution is analytically intractable, we sample from it using the Markov Chain Monte Carlo (MCMC) technique \cite{brooks2011handbook}, 
which samples the parameter space with a frequency proportional to the desired posterior $p(\boldsymbol\theta|\mathcal D)$ (See %\ref{SupSec:MCMC} 
``Markov Chain Monte Carlo sampling'' section in the supplementary material). 

% As a result, instead of obtaining a single value as a point estimate for prediction of the response variable, a full distribution is recovered. 
As a result, instead of obtaining a single value as the prediction for the response variable, the ensemble model produces a full distribution that takes into account the uncertainty in model parameters. More precisely, for a new input $\mathbf{x}^*$ (not present in $\mathcal D$), the ensemble model $F$ provides the probability that the response is $y$, when trained with data $\mathcal D$ (i.e. the full predictive posterior distribution):
\begin{equation}\label{Eq:post_pred}
p(y| \mathbf x^*,\mathcal D)=\int p(y| \mathbf x^*,\boldsymbol\theta)p(\boldsymbol\theta|\mathcal D)\mathrm d\boldsymbol\theta = \int \mathcal N(y; \boldsymbol{w}^T \mathbf f,\sigma^2) p(\boldsymbol\theta|\mathcal D) \mathrm d\boldsymbol\theta.
\end{equation}
 where $p(y| \mathbf x^*,\boldsymbol\theta)$ is the predictive distribution of $y$ given input $\mathbf x^*$ and model parameters $\boldsymbol\theta$, $p(\boldsymbol\theta|\mathcal D)$ is the posterior distribution of model parameters given data $\mathcal D$, and $\mathbf f\equiv\mathbf f(\mathbf x^*)$ for the sake of clarity.

\subsubsection{Optimization: suggesting next steps}
The optimization phase leverages the predictive model described in the previous section to find inputs that are predicted to bring us closer to our objective (i.e. maximize or minimize response, or achieve a desired response level). In mathematical terms, we are looking for a set of $N_r$ suggested inputs $\mathbf x_r \in \mathcal X; r=1,\dots,N_r$, that optimize the response with respect to the desired objective. Specifically, we want a process that: 
\begin{itemize}
\item[i)] optimizes the predicted levels of the response variable; 
\item[ii)] can explore the regions of input phase space associated with high uncertainty in predicting response, if desired; 
\item[iii)] provides a set of different recommendations, rather than only one. 
\end{itemize}

In order to meet these three requirements, we define the optimization problem formally as
\begin{eqnarray*}
    \arg\max_{\mathbf x}G(\mathbf x) \\
    \text{s.t. } \mathbf x \in \mathcal B 
\end{eqnarray*}
where the surrogate function $G(\mathbf x)$ is defined as:
\begin{equation} \label{Eq:G_function}
G(\mathbf x) = \left\{ \begin{array}{lc} 
  \,\,\,\,\,(1-\alpha)\mathbb E(y)+\alpha \mathrm{Var}(y)^{1/2} & \text{(maximization case)} \\
  -(1-\alpha)\mathbb E(y)+\alpha \mathrm{Var}(y)^{1/2} & \text{(minimization case)} \\
  -(1-\alpha)||\mathbb E(y)-y^*||_2^2+\alpha \mathrm{Var}(y)^{1/2} & \text{(specification case)}
       \end{array} \right.
\end{equation}
depending on which mode \textsf{ART} is operating in (see \ref{Sec:Key_cap}``Key capabilities'' section). Here, $y^*$ is the target value for the response variable, $y=y(\mathbf x)$, $\mathbb E(y)$ and $\mathrm{Var}(y)$ denote the expected value and variance respectively (see ``Expected value and variance for ensemble model'' in the supplementary material), $||\mathbf x||_2^2  = \sum_i x_i^2$ denotes Euclidean distance, and the parameter $\alpha \in [0,1]$ represents the exploitation-exploration trade-off (see below). The constraint $\mathbf x \in \mathcal B$ characterizes the lower and upper bounds for each input feature (e.g. protein levels cannot increase beyond a given, physical, limit). These bounds can be provided by the user (see details in the %\ref{Sec:Implementation}
``Implementation'' section in the supplementary material); otherwise default values are computed from the input data as described in the ``Input space set $\mathcal{B}$'' section in the supplementary material.

Requirements i) and ii) are both addressed by borrowing an idea from Bayesian optimization\cite{Mockus1989}: optimization of a parametrized surrogate function which accounts for both exploitation and exploration. Namely, our objective function $G(\mathbf x)$ takes the form of the upper confidence bound \cite{Snoek2012} given in terms of a weighted sum of the expected value and the variance of the response (parametrized by $\alpha$, Eq.\ \ref{Eq:G_function}). This scheme accounts for both exploitation and exploration: for the maximization case, for example, for $\alpha=1$ we get $G(\mathbf x) = \mathrm{Var}(y)^{1/2}$, so the algorithm suggests next steps that maximize the response variance, thus \emph{exploring} parts of the phase space where our model shows high predictive uncertainty. For $\alpha=0$, we get $G(\mathbf x) = E(y)$, and the algorithm suggests next steps that maximize the expected response, thus \emph{exploiting} our model to obtain the best response. Intermediate values of $\alpha$ produce a mix of both behaviors. We recommend setting $\alpha$ to values slightly smaller than one for early-stage DBTL cycles, thus allowing for more systematic exploration of the space so as to build a more accurate predictive model in the subsequent DBTL cycles. If the objective is purely to optimize the response, we recommend setting $\alpha=0$.

\begin{figure}[h!]
\begin{center}
\includegraphics[width=1\columnwidth]{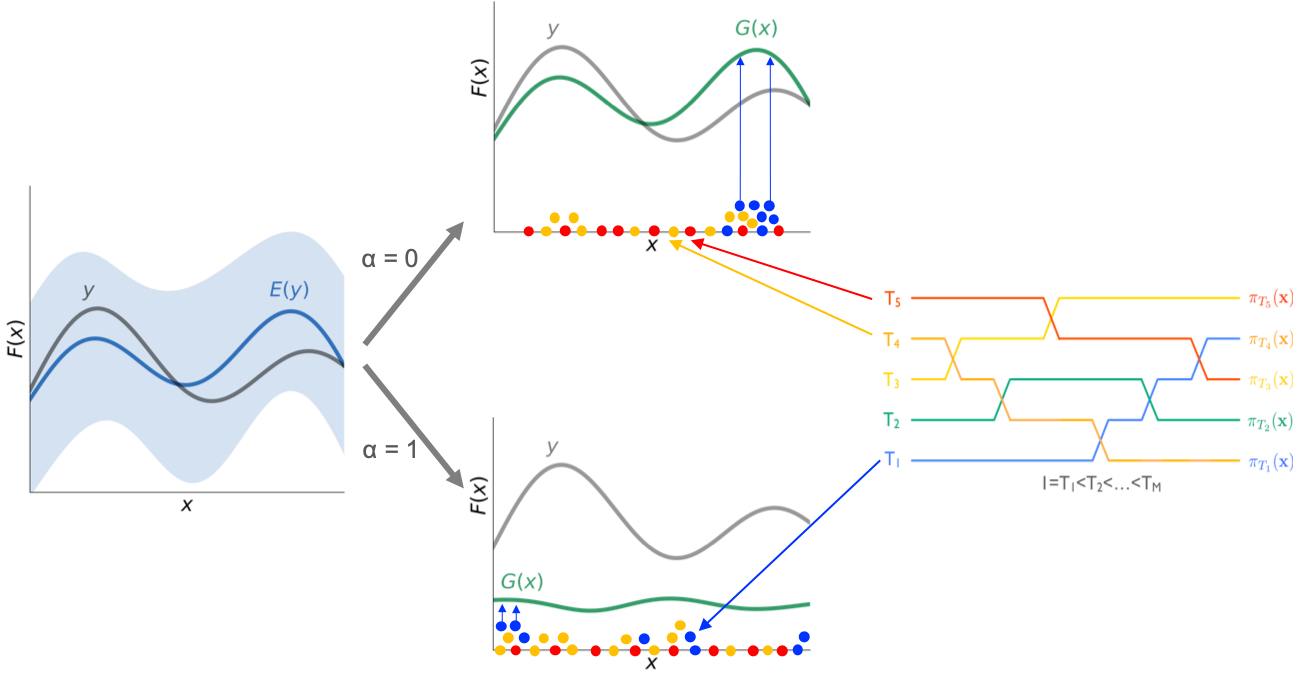}
\end{center}
\caption{\textbf{\textsf{ART} chooses recommendations for next steps by sampling the modes of a surrogate function}. The leftmost panel shows the true response $y$ (e.g.\ biofuel production to be optimized) as a function of the input $\mathbf x$ (e.g.\ proteomics data), as well as the expected response $E(y)$ after several DBTL cycles, and its 95\% confidence interval (blue). Depending on whether we prefer to \emph{explore} the phase space where the model is least accurate or \emph{exploit} the predictive model to obtain the highest possible predicted responses, we will seek to optimize a surrogate function $G(\mathbf x)$ (Eq.~\ref{Eq:G_function}), where the exploitation-exploration parameter is $\alpha = 0$ (pure exploitation), $\alpha = 1$ (pure exploration) or anything in between. Parallel-Tempering-based MCMC sampling (center and right side) produces sets of vectors $\mathbf x$ (colored dots) for different ``temperatures'': higher temperatures (red) explore the full phase space, while lower temperature chains (blue) concentrate in the nodes (optima) of $G(\mathbf x)$. Exchange between different ``temperatures'' provides more efficient sampling without getting trapped in local optima. Final recommendations (blue arrows) to improve response are provided from the lowest temperature chain, and chosen such that they are not too close to each other and to experimental data (at least 20\% difference).}
\label{fig:recommendations}
\end{figure}

In order to address (iii), as well as to avoid entrapment in local optima and search the phase space more effectively, we choose to solve the optimization problem through sampling. More specifically, we draw samples from a target distribution defined as
\begin{equation}\label{Eq:target_distr}
    \pi(\mathbf x)\propto \exp(G(\mathbf x))p(\mathbf x),
\end{equation}
where $p(\mathbf x)= \mathcal U(\mathcal B)$ can be interpreted as the uniform `prior' on the set $\mathcal B$, and $\exp(G(\mathbf x))$ as the `likelihood' term of the target distribution. Sampling from $\pi$ implies optimization of the function $G$ (but not reversely), since the modes of the distribution $\pi$ correspond to the optima of $G$. As we did before, we resort to MCMC for sampling. The target distribution is not necessarily differentiable and may well be complex. For example, if it displays more than one mode, as is often the case in practice, there is a risk that a Markov chain gets trapped in one of them. In order to make the chain explore all areas of high probability one can ``flatten/melt down'' the roughness of the distribution by tempering. For this purpose, we use the Parallel Tempering algorithm \cite{Earl2005} for optimization of the objective function through sampling, in which multiple chains at different temperatures are used for exploration of the target distribution (Fig.~\ref{fig:recommendations}).

% Acceptance criterion
% $$\alpha=\min\left\{ 1, \frac{\pi(\mathbf x^{'})}{\pi(\mathbf x)}\right\}$$
% ensures that a proposed sample with objective function value higher than the current sample is always accepted and proposals with lower values are accepted with probability that increases if the difference in objective values is smaller.

%How does Parallel Tempering work: $k$ chains run in parallel and draw samples from distributions $\pi\propto ,k=1,\dots,K$... Occasionally, the chains exchange states, enabling ...

\medskip

\emph{Choosing recommendations for the next cycle}

 After drawing a certain number of samples from $\pi(\mathbf x)$ we need to choose recommendations for the next cycle,  making sure that they are sufficiently different from each other as well as from the input experimental data. To do so, first we find a sample with optimal $G(\mathbf x)$ (note that $G(\mathbf x)$ values are already calculated and stored). We only accept this sample as a recommendation if there is \emph{at least one} feature whose value is different by at least a factor $\gamma$ (e.g.\ 20\% difference, $\gamma =0.2$) from the values of that feature in \emph{all} data points ${\mathbf x}\in \mathcal D$. Otherwise, we find the next optimal sample and check the same condition. This procedure is repeated until the desired number of recommendations are collected, and the condition involving $\gamma$ is satisfied for all previously collected recommendations and all data points. In case all draws are exhausted without collecting the sufficient number of recommendations, we decrease the factor $\gamma$ and repeat the procedure from the beginning. Pseudo code for this algorithm can be found in Algorithm 1 %\ref{Alg:Recommendations}
 in the supplementary material. The probability of success for these recommendations is computed as indicated in the ``Success probability calculation'' section in the supplementary material.

\subsection{Implementation}\label{Sec:Implementation}

\textsf{ART} is implemented Python 3.6 and should be used under this version (see below for software availability).
%{\color{blue}(The source code can be downloaded from \url{https://github.com/JBEI/AutomatedRecommendationTool} and installed by the command \texttt{python setup.py install}.)} 
Figure S1 %\ref{fig:codeStructure} 
represents the main code structure and its dependencies to external packages. In the %\ref{SuplSec:Implementation}
``Implementation'' section of the supplementary material, we provide an explanation for the main modules and their functions.

\section{Results and discussion}

\subsection{Using simulated data to test ART}

\begin{figure}[h!]
\begin{center}
\includegraphics[width=1\columnwidth]{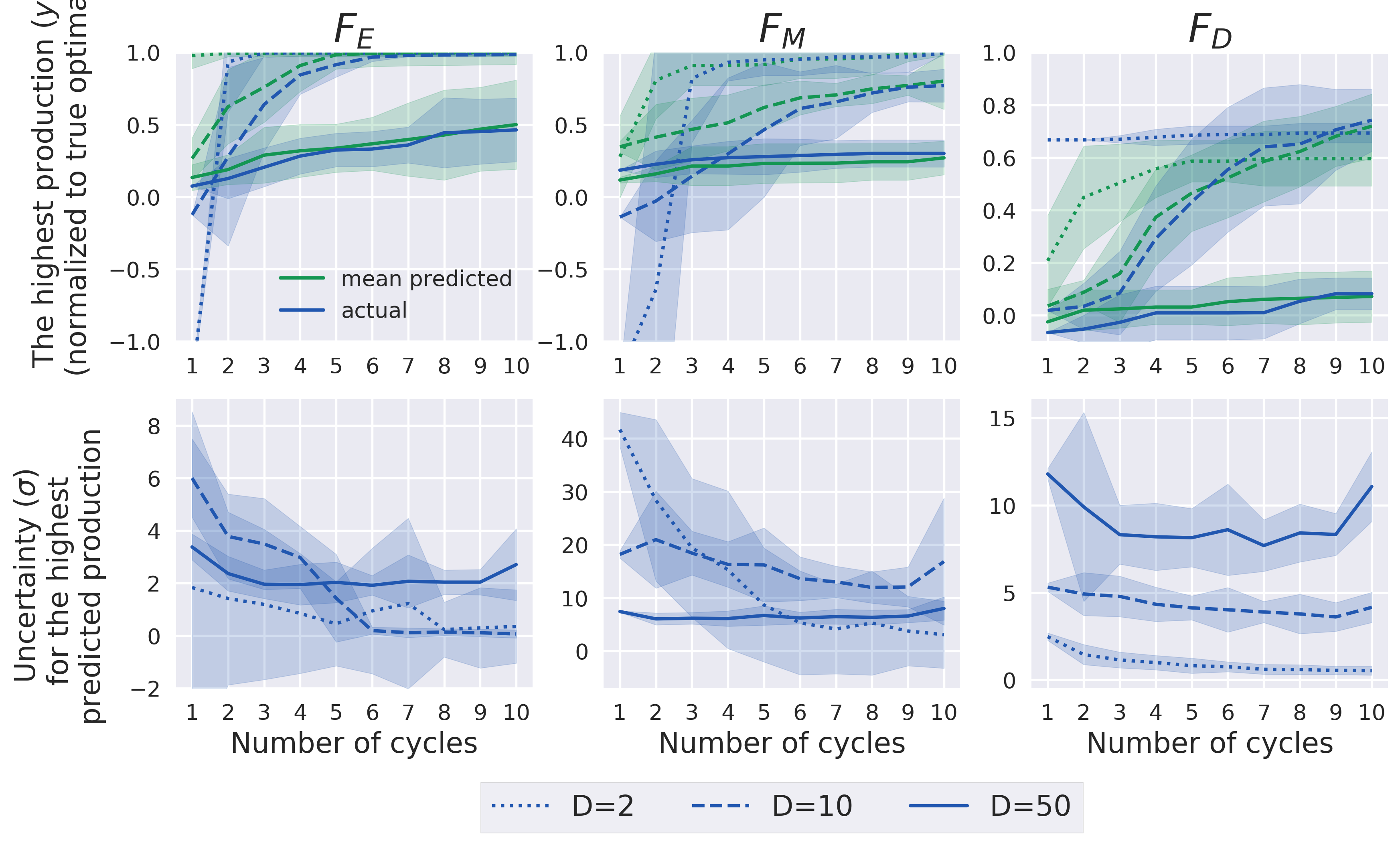}
\end{center}
\caption{\textbf{\textsf{ART} performance improves significantly by proceeding beyond the usual two Design-Build-Test-Learn cycles.} Here we show the results of testing \textsf{ART}'s performance with synthetic data obtained from functions of different levels of complexity (see Table~\ref{Tab:functions}), different phase space dimensions (2, 10 and 50), and different amounts of training data (DBTL cycles). The top row presents the results of the simulated metabolic engineering in terms of highest production achieved so far for each cycle (as well as the corresponding \textsf{ART} predictions). The production increases monotonically with a rate that decreases as the problem is harder to learn, and the dimensionality increases.  The bottom row shows the uncertainty in \textsf{ART}'s production prediction, given by the standard deviation of the response distribution (Eq.~\ref{Eq:post_pred}). This uncertainty decreases markedly with the number of DBTL cycles, except for the highest number of dimensions. In each plot, lines and shaded areas represent the estimated mean values and 95\% confidence intervals, respectively, over 10 repeated runs. Mean Absolute Error (MAE) and training and test set definitions can be found in Fig.~S2%\ref{FigMAE}
.}
\label{fig:F_E}
\end{figure}

Synthetic data sets allow us to test how \textsf{ART} performs when confronted by problems of different difficulty and dimensionality, as well as gauge the effect of the availability of more training data. In this case, we tested the performance of \textsf{ART} for 1--10 DBTL cycles, three problems of increasing difficulty ($F_E$, $F_M$ and $F_D$, see Table~\ref{Tab:functions}), and three different dimensions of input space ($D=2, 10$ and $50$, Fig.~\ref{fig:F_E}). %We simulated the DBTL processes by starting with a training set given by 16 strains and measurements (mimicking the 48 wells of throughput of a typical automated fermentation platform\cite{unthan2015bioprocess}). We limited ourselves to the maximization case, and at each DBTL cycle generated 16 recommendations that maximize the objective function given by Eq.~\eqref{Eq:G_function}. 
We simulated the DBTL processes by starting with a training set given by 16 strains (Latin Hypercube\cite{McKay1979} draws) and the associated measurements (from Table~\ref{Tab:functions} functions). We limited ourselves to the maximization case, and at each DBTL cycle, generated 16 recommendations that maximize the objective function given by Eq.~\eqref{Eq:G_function}. This choice mimicked triplicate experiments in the 48 wells of throughput of a typical automated fermentation platform\cite{unthan2015bioprocess}.
We employed a \emph{tempering} strategy for the exploitation-exploration parameter, i.e.\ assign $\alpha=0.9$ at start for an exploratory optimization, and gradually decreased the value to $\alpha=0$ in the final DBTL cycle for the exploitative maximization of the production levels.

\begin{table}[!h]
\centering
\renewcommand{\arraystretch}{1.8}% Spread rows out...
%   \begin{tabular}{| c |>{\centering}c |>{\centering\arraybackslash}m{1in}|}
  \begin{tabular}{| c| >{\centering}c | >{\centering\arraybackslash}m{4cm}|}
    \hline
	 \textbf{Easy}   & $\begin{array} {c} F_E(\mathbf x) = \\ -\frac{1}{d}\sum_i^d (x_i-5)^2 + \exp\left(-\sum_i x_i^2\right) + 25 \end{array}$ & \includegraphics[width=0.21\columnwidth]{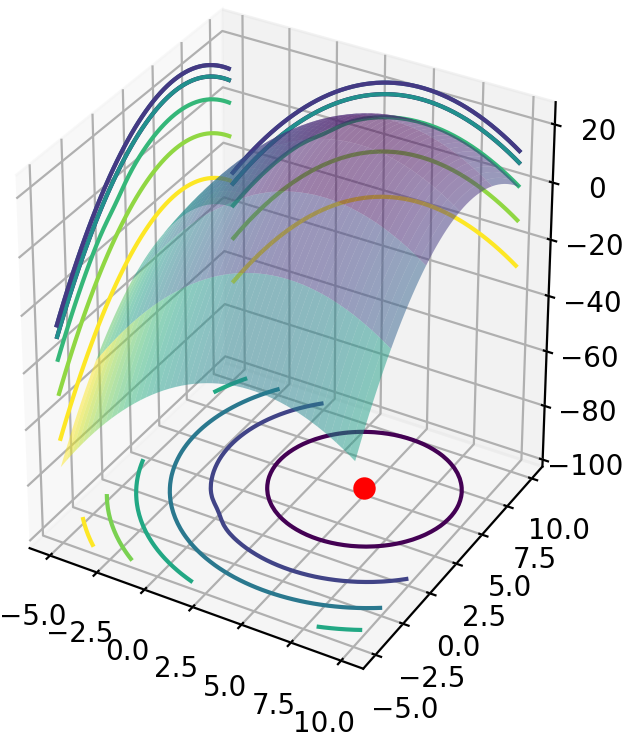}  \\ \hline
     \textbf{Medium} & $\begin{array} {c} F_M(\mathbf x) = \\ \frac{1}{d}\sum_i^d \left(x_i^4 - 16x_i^2 +5x_i\right) \end{array}$ & \includegraphics[width=0.21\columnwidth]{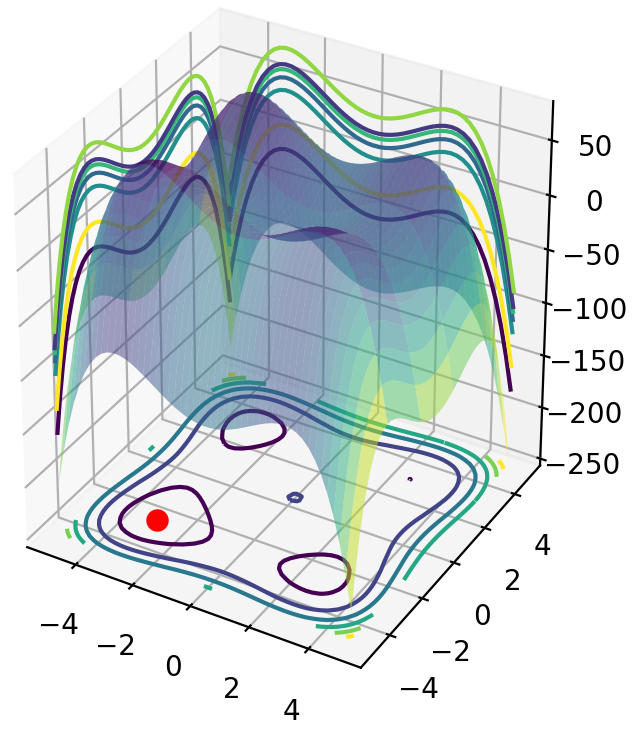}  \\ \hline
	 \textbf{Difficult}   & $\begin{array} {c} F_D(\mathbf x) = \\ \sum_i^d \sqrt{x_i}\sin(x_i) \end{array}$ & \includegraphics[width=0.21\columnwidth]{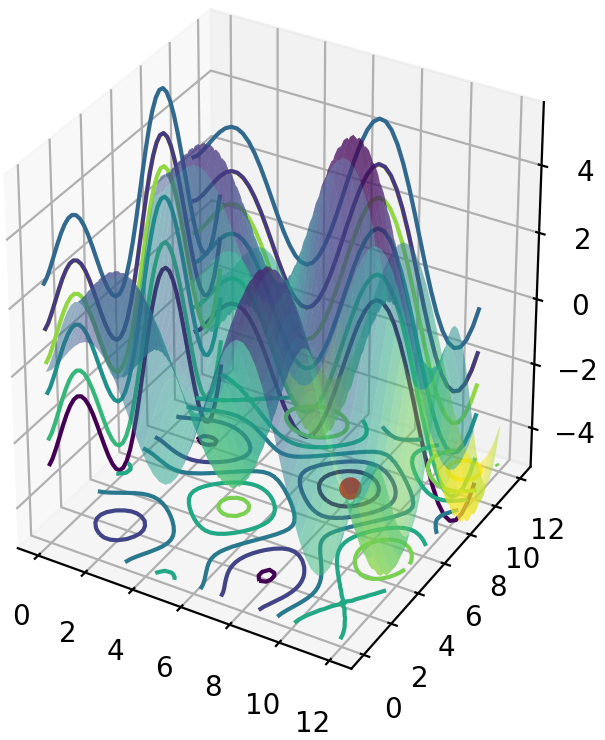}  \\ \hline
	\end{tabular}
	\caption{Functions presenting different levels of difficulty to being learnt, used to produce synthetic data and test \textsf{ART}'s performance (Fig.~\ref{fig:F_E}).}
	\label{Tab:functions}
\end{table}

\textsf{ART} performance improves significantly as more data are accrued with additional DTBL cycles. Whereas the prediction error, given in terms of Mean Average Error (MAE), remains constantly low for the training set (i.e.\ \textsf{ART} is always able to reliably predict data it has already seen), the MAE for the test data (data \textsf{ART} has not seen) in general decreases markedly only with the addition of more DBTL cycles (Fig.~S2%\ref{FigMAE}
). The exceptions are the most complicated problems: those exhibiting highest dimensionality ($D=50$), where MAE stays approximately constant, and the difficult function $F_D$, which exhibits a slower decrease. 
Furthermore, the best production among the 16 recommendations obtained in the simulated process %, given in terms of the highest mean predicted production,
increases monotonically with more DBTL cycles: faster for easier problems and lower dimensions and more slowly for harder problems and higher dimensions. Finally, the uncertainty in those predictions decreases as more DBTL cycles proceed (Fig.~\ref{fig:F_E}). Hence, more data (DBTL cycles) almost always translates into better predictions and production. However, we see that these benefits are rarely reaped with only the 2 DBTL cycles customarily used in metabolic engineering (see examples in the next sections): \textsf{ART} (and ML in general) becomes only truly efficient when using 5--10 DBTL cycles. 

Different experimental problems involve different levels of difficulty when being learnt  (i.e.\ being predicted accurately), and this can only be assessed empirically. Low dimensional problems can be easily learnt, whereas exploring and learning a {\color{black}50-dimensional} landscape is very slow (Fig.~\ref{fig:F_E}). Difficult problems (i.e.\ less monotonic landscapes) take more data to learn and traverse than easier ones. We will showcase this point in terms of real experimental data when comparing the biofuel project (easy) versus the dodecanol project (hard) below. However, it is {\color{black}(difficult)} to decide a priori whether a given real data project or problem will be easy or hard to learn---the only way to determine this is by checking the improvements in prediction accuracy as more data is added. In any case, a starting point of at least $\sim 100$ instances is highly recommendable to obtain proper statistics.

\subsection{Improving the production of renewable biofuel}

The optimization of the production of the renewable biofuel limonene through synthetic biology will be our first demonstration of \textsf{ART} using real-life experimental data. Renewable biofuels are almost carbon neutral because they only release into the atmosphere the carbon dioxide that was taken up in growing the plant biomass they are produced from. Biofuels from renewable biomass have been estimated to be able to displace $\sim$30\% of petroleum consumption~\cite{langholtz20162016} and are seen as the most viable option for decarbonizing sectors that are challenging to electrify, such as heavy-duty freight and aviation\cite{renouard2010improving}.

Limonene is a molecule that can be chemically converted to several pharmaceutical and commodity chemicals~\cite{keasling2010manufacturing}. If hydrogenated, for example, it has low freezing point and is immiscible with water, characteristics which are ideal for next generation jet-biofuels and fuel additives that enhance cold weather performance~\cite{tracy2009hydrogenated,ryder2009jet}. Limonene has been traditionally obtained from plant biomass, as a byproduct of orange juice production, but fluctuations in availability, scale and cost limit its use as biofuel\cite{duetz2003biotransformation}. The insertion of the plant genes responsible for the synthesis of limonene in a host organism (e.g.\ a bacteria), however, offers a scalable and cheaper alternative through synthetic biology. Limonene has been produced in  \emph{E.\ coli} through an expansion of the celebrated mevalonate pathway (Fig.\ 1a in \citet{alonso2013metabolic}), used to produce the antimalarial precursor artemisinin\cite{paddon2013high} and the biofuel farnesene\cite{meadows2016rewriting}, and which forms the technological base on which the company Amyris was founded (valued $\sim$\$300M ca.\ 2019). This version of the mevalonate pathway is composed of seven genes obtained from such different organisms as  \emph{S. cerevesiae}, \emph{S. aureus}, and \emph{E. coli}, to which two genes have been added: a geranyl-diphosphate synthase and a limonene synthase obtained from the plants \emph{A. grandis} and \emph{M. spicata}, respectively. 

For this demonstration, we use historical data from \citet{Alonso-Gutierrez2015}, where 27 different variants of the pathway (using different promoters, induction times and induction strengths) were built. Data collected for each variant involved limonene production and protein expression for each of the nine proteins involved in the synthetic pathway. These data were used to feed Principal Component Analysis of Proteomics (PCAP) \cite{Alonso-Gutierrez2015}, an algorithm using principal component analysis to suggest new pathway designs. The PCAP recommendations, used to engineer new strains, resulted in a 40\% increase in production for limonene, and 200\% for bisabolene (a molecule obtained from the same base pathway). This small amount of available instances (27) to train the algorithms is typical of synthetic biology/metabolic engineering projects. Although we expect automation to change the picture in the future\cite{Carbonell2019}, the lack of large amounts of data has determined our machine learning approach in \textsf{ART} (i.e.\ no deep neural networks).

\textsf{ART} is able to not only recapitulate the successful predictions obtained by PCAP improving limonene production, but also provides a systematic way to obtain them as well as the corresponding uncertainty. In this case, the training data for \textsf{ART} are the concentrations for each of the nine proteins in the heterologous pathway (input), and the production of limonene (response). The objective is to maximize limonene production. We have data for two DBTL cycles, and we use \textsf{ART} to explore what would have happened if we have used \textsf{ART} instead of PCAP for this project.

\begin{figure}[h!]
\begin{center}
\includegraphics[width=1.\columnwidth]{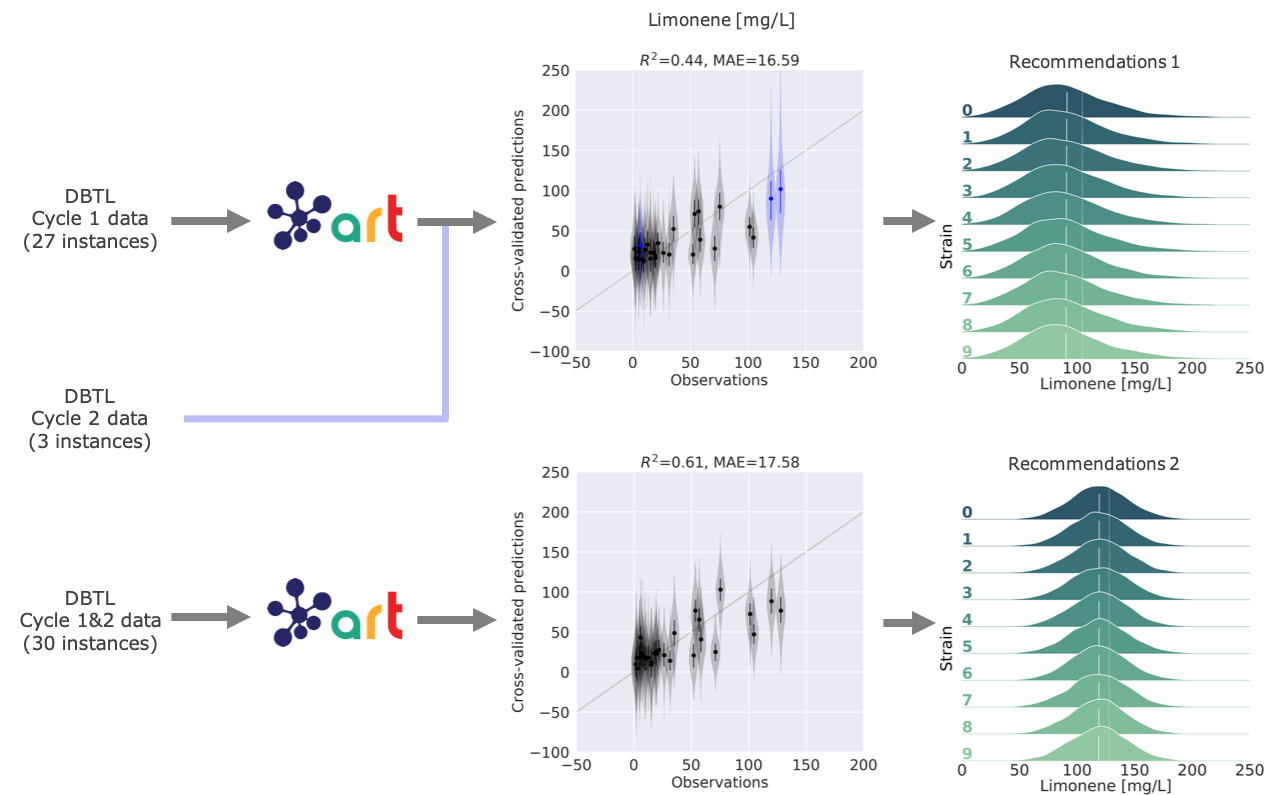}
\end{center}
\caption{\textbf{\textsf{ART} provides effective recommendations to improve renewable biofuel (limonene) production}. We used the first DBTL cycle data (27 strains, top) to train \textsf{ART} and recommend new protein targets (top right). The \textsf{ART} recommendations were very similar to the protein profiles that eventually led to a 40\% increase in production (Fig.\ \ref{fig:MLcomp}). \textsf{ART} predicts mean production levels for the second DBTL cycle strains which are very close to the experimentally measured values (three blue points in top graph). Adding those three points from DBTL cycle 2 provides a total of 30 strains for training that lead to recommendations predicted to exhibit higher production and narrower distributions (bottom right). Uncertainty for predictions is shown as probability distributions for recommendations and violin plots for the cross-validated predictions. $R^2$ and Mean Absolute Error (MAE) values are only for cross-validated mean predictions (black data points).}
\label{fig:example1}
\end{figure}

\begin{figure}[!h]
\begin{center}
\includegraphics[width=1.\columnwidth]{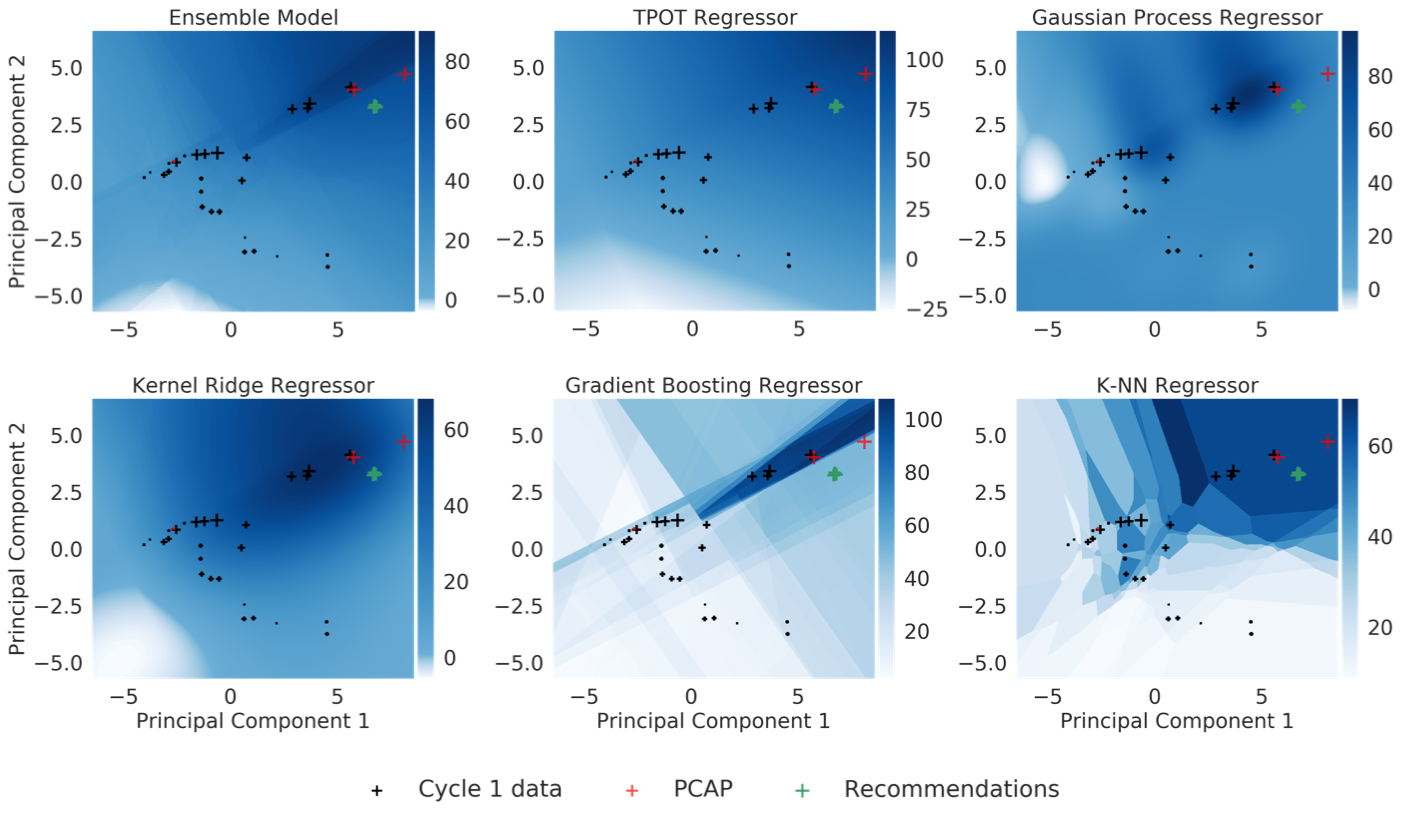}
\end{center}
\caption{\textbf{All machine learning algorithms point in the same direction to improve limonene production, in spite of quantitative differences in prediction}. Cross sizes indicate experimentally measured limonene production in the proteomics phase space (first two principal components shown from principal component analysis, PCA). The color heatmap indicates the limonene production predicted by a set of base regressors and the final ensemble model (top left) that leverages all the models and conforms the base algorithm used by \textsf{ART}. Although the models differ significantly in the actual quantitative predictions of production, the same qualitative trends can be seen in all models (i.e.\ explore upper right quadrant for higher production), justifying the ensemble approach used by \textsf{ART}. The \textsf{ART} recommendations (green) are very close to the PCAP recommendations (red) that were experimentally tested to improve production by 40\%.}
\label{fig:MLcomp}
\end{figure}

We used the data from DBLT cycle 1 to train \textsf{ART} and recommend new strain designs (i.e.\ protein profiles for the pathway genes, Fig.~\ref{fig:example1}). The model trained with the initial 27 instances provided reasonable cross-validated predictions for production of this set ($R^2=0.44$), as well as the three strains which were created for DBTL cycle 2 at the behest of PCAP (Fig. \ref{fig:example1}). This suggests that \textsf{ART} would have easily recapitulated the PCAP results. Indeed, the \textsf{ART} recommendations are very close to the PCAP recommendations (Fig.~\ref{fig:MLcomp}). 
%While we cannot experimentally test these recommendations, \textsf{ART}'s capability to predict the outcome of the PCAP suggestions (blue points in Fig.\ \ref{fig:example1}) suggests that \textsf{ART}'s results would have improved upon those of PCAP. 
Interestingly, we see that while the quantitative predictions of each of the individual models were not very accurate, they all signaled towards the same direction in order to improve production, hence showing the importance of the ensemble approach (Fig.~\ref{fig:MLcomp}).

Training \textsf{ART} with experimental results from DBTL cycles 1 and 2 results in even better predictions ($R^2=0.61$), highlighting the importance of the availability of large amounts of data to train ML models. This new model suggests new sets of strains predicted to produce even higher amounts of limonene. Importantly, the uncertainty in predicted production levels is significantly reduced with the additional data points from cycle 2.

%\clearpage

\subsection{Brewing hoppy beer without hops by bioengineering yeast}
Our second example involves bioengineering yeast (\emph{S. cerevisiae}) to produce hoppy beer without the need for hops\cite{denby2018industrial}. To this end, the ethanol-producing yeast used to brew the beer, was modified to also synthesize the metabolites linalool  (L) and geraniol (G), which impart hoppy flavor (Fig.\ 2B in \citet{denby2018industrial}). Synthesizing linalool and geraniol through synthetic biology is economically advantageous because growing hops is water and energetically intensive, and their taste is  {\color{black}highly} variable from crop to crop. Indeed, a startup (Berkeley Brewing Science\cite{bbs}) was generated from this technology.

\begin{figure}[h!]
\begin{center}
\includegraphics[width=1\columnwidth]{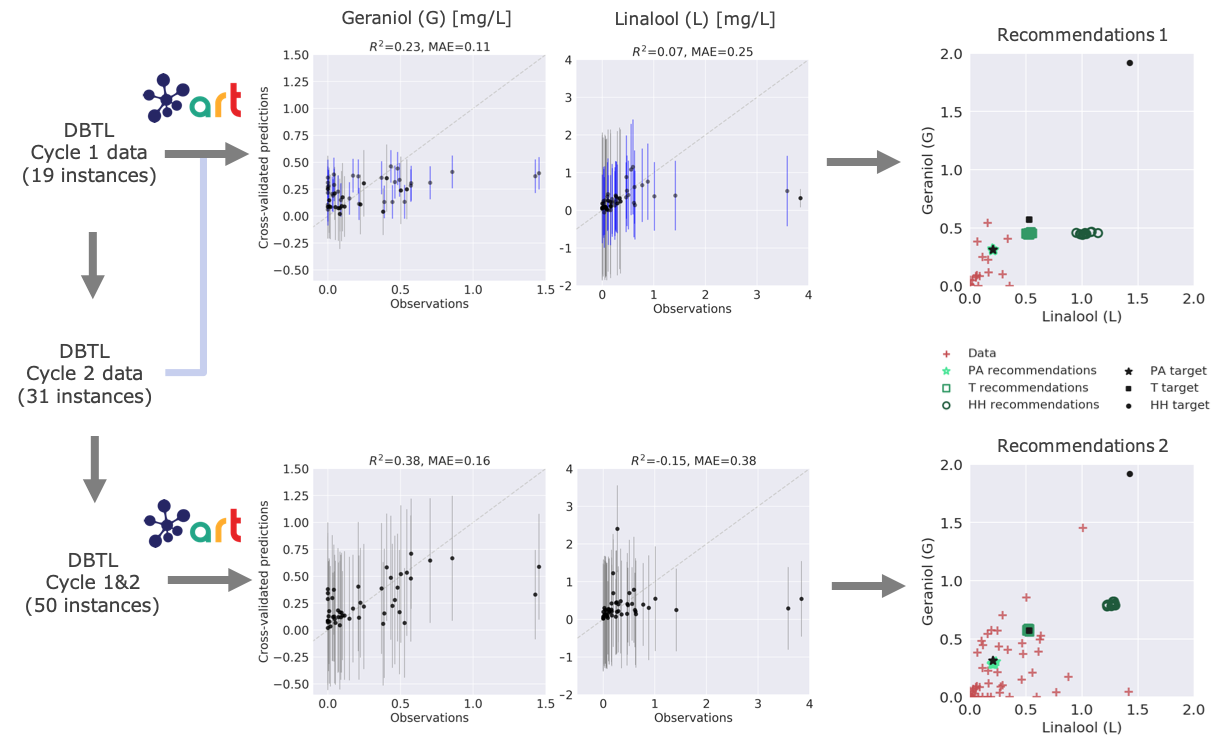}
\end{center}
\caption{\textbf{\textsf{ART} produces effective recommendations to bioengineer yeast to produce hoppy beer without hops.} The 19 instances in the first DBTL cycle were used to train \textsf{ART}, but it did not show an impressive predictive power (particularly for L, top middle). In spite of it, \textsf{ART} is still able to recommend protein profiles predicted to reach the Pale Ale (PA) target flavor profile, and others which were close to the Torpedo (T) metabolite profile (top right, green points showing mean predictions). Adding the 31 strains for the second DBTL cycle improves predictions for G but not for L (bottom). The expanded range of values for G \& L provided by cycle 2 allows \textsf{ART} to recommend profiles which are predicted to reach targets for both beers (bottom right), but not Hop Hunter (HH). Hop Hunter displays a very different metabolite profile from the other beers, well beyond the range of experimentally explored values of G \& L, making it impossible for \textsf{ART} to extrapolate that far. Notice that none of the experimental data (red crosses) matched exactly the desired targets (black symbols), but the closest ones were considered acceptable. $R^2$ and Mean Absolute Error (MAE) values are for cross-validated mean predictions (black data points) only. Bars indicate 95\% credible interval of the predictive posterior distribution.}
\label{fig:example2}
\end{figure}

\textsf{ART} is able to efficiently provide the proteins-to-production mapping that required three different types of mathematical models in the original publication, paving the way for a systematic approach to beer flavor design. The challenge is different in this case as compared to the previous example (limonene): instead of trying to maximize production, the goal is to reach a particular level of linalool and geraniol so as to match a known beer tasting profile (e.g.\ Pale Ale, Torpedo or Hop Hunter, Fig.~\ref{fig:example2}). \textsf{ART} can provide this type of recommendations, as well. For this case, the inputs are the expression levels for the four different proteins involved in the pathway, and the response are the concentrations of the two target molecules (L and G), for which we have desired targets. We have data for two DBTL cycles involving 50 different strains/instances (19 instances for the first DBTL cycle and 31 for the second one, Fig.~\ref{fig:example2}). As in the previous case, we use this data to simulate the outcomes we would have obtained in case \textsf{ART} had been available for this project. 

The first DBTL cycle provides a very limited number of 19 instances to train \textsf{ART}, which performs passably on this training set, and poorly on the test set provided by the 31 instances from DBTL cycle 2 (Fig.~\ref{fig:example2}). Despite this small amount of training data, the model trained in DBTL cycle 1 is able to recommend new protein profiles that are predicted to reach the Pale Ale target (Fig.~\ref{fig:example2}). Similarly, this DBTL cycle 1 model was almost able to reach (in predictions) the L and G levels for the Torpedo beer, which will be finally achieved in DBTL cycle 2 recommendations, once more training data is available. For the Hop Hunter beer, recommendations from this model were not close to the target.     

The model for the second DBTL cycle leverages the full 50 instances from cycles 1 and 2 for training and is able to provide recommendations predicted to attain two out of three targets. The Pale Ale target L and G levels were already predicted to be matched in the first cycle; the new recommendations are able to maintain this beer profile. The Torpedo target was almost achieved in the first cycle, and is predicted to be reached in the second cycle recommendations. Finally, Hop Hunter  target L and G levels are very different from the other beers and cycle 1 results, so neither cycle 1 or 2 recommendations can predict protein inputs achieving this taste profile. \textsf{ART} has only seen two instances of high levels of L and G and cannot extrapolate well into that part of the metabolic phase space. \textsf{ART}'s exploration mode, however, can suggest experiments to explore this space.

Quantifying the prediction uncertainty is of fundamental importance to gauge the reliability of the recommendations, and the full process through several DBTL cycles. In the end, the fact that \textsf{ART} was able to recommend protein profiles predicted to match the Pale Ale and Torpedo taste profiles only indicates that the optimization step (see "Optimization: suggesting next steps" section) works well. The actual recommendations, however, are only as good as the predictive model. In this regard, the predictions for L and G levels shown in Fig.~\ref{fig:example2} (right side) may seem deceptively accurate, since they are only showing the average predicted production. Examining the full probability distribution provided by \textsf{ART} shows a very broad spread for the L and G predictions (much broader for L than G, Fig.~S3%\ref{FigHopless_recom_with_distr}
). These broad spreads indicate that the model still has not converged and that recommendations will probably change significantly with new data. Indeed, the protein profile recommendations for the Pale Ale changed markedly from DBTL cycle 1 to 2, although the average metabolite predictions did not (left panel of Fig.~S4%\ref{FigHoplessPCAs}
). All in all, these considerations indicate that quantifying the uncertainty of the predictions is important to foresee the smoothness of the optimization process.

At any rate, despite the limited predictive power afforded by the cycle 1 data, \textsf{ART} recommendations guide metabolic engineering effectively. For both of the Pale Ale and Torpedo cases, \textsf{ART} recommends exploring parts of the proteomics phase space such that the final protein profiles (that were deemed close enough to the targets), lie between the first cycle data and these recommendations (Fig.~S4%\ref{FigHoplessPCAs}
). Finding the final target becomes, then, an interpolation problem, which is much easier to solve than an extrapolation one. These recommendations improve as \textsf{ART} becomes more accurate with more DBTL cycles.

\subsection{Improving dodecanol production}

The final example is one of a failure (or at least a mitigated success), from which as much can be learnt as from the previous successes. 
\citet{Opgenorth2019} used machine learning to drive two DBTL cycles to improve production of 1-dodecanol in \emph{E. coli}, a medium-chain fatty acid used in detergents, emulsifiers, lubricants and cosmetics. This example illustrates the case in which the assumptions underlying this metabolic engineering and modeling approach (mapping proteomics data to production) fail. Although a $\sim$20\% production increase was achieved, the machine learning algorithms were not able to produce accurate predictions with the low amount of data available for training, and the tools available to reach the desired target protein levels were not accurate enough. 

This project consisted of two DBTL cycles comprising 33 and 21 strains, respectively, for three alternative pathway designs (Fig.\ 1 in \citet{Opgenorth2019}, Table S4% \ref{TabDodecanol_data}
). The use of replicates increased the number of instances available for training to 116 and 69 for cycle 1 and 2, respectively. The goal was to modulate the protein expression by choosing Ribosome Binding Sites (RBSs, the mRNA sites to which ribosomes bind in order to translate proteins) of different strengths for each of the three pathways. The idea was for the machine learning to operate on a small number of variables ($\sim$3 RBSs) that, at the same time, provided significant control over the pathway. As in previous cases, we will show how \textsf{ART} could have been used in this project. The input for \textsf{ART} in this case consists of the concentrations for each of three proteins (different for each of the three pathways), and the goal was to maximize 1-dodecanol production. 

\begin{figure}[h!]
\begin{center}
\includegraphics[width=1\columnwidth]{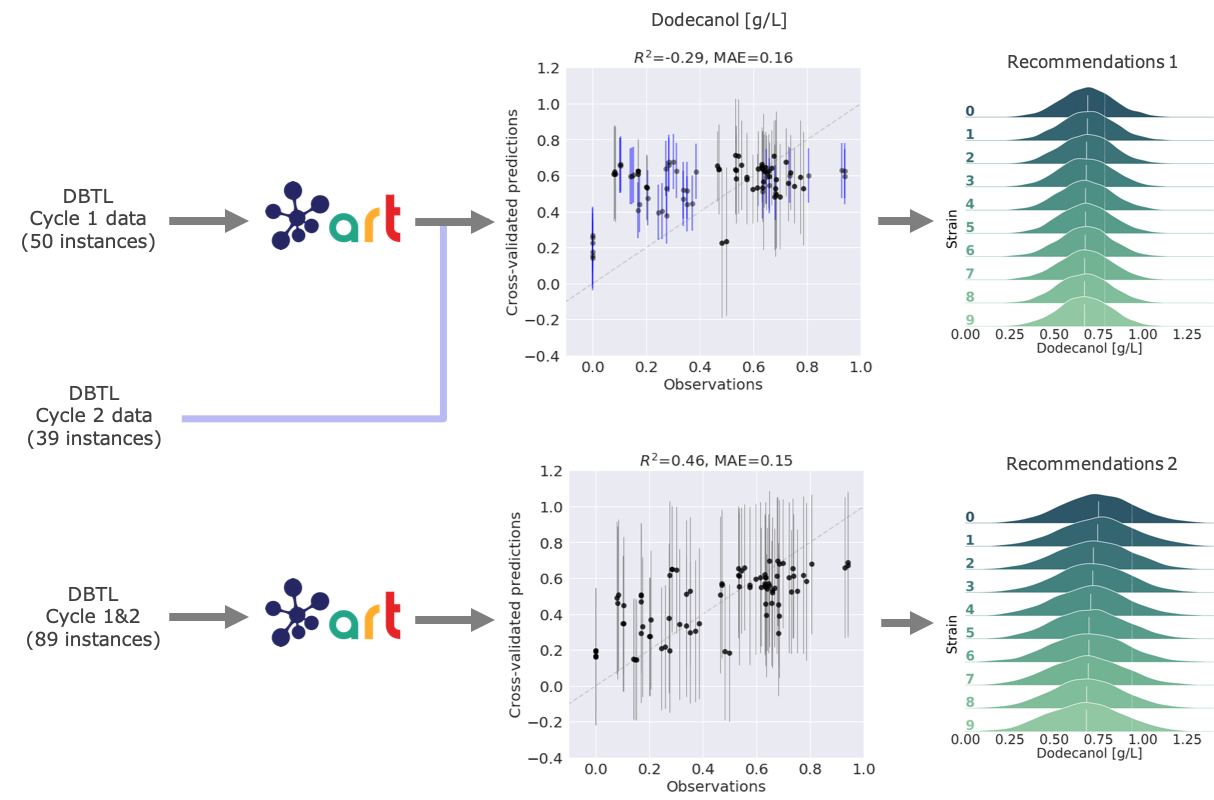}
\end{center}
\caption{\textbf{\textsf{ART}'s predictive power is heavily compromised in the dodecanol production example}. Although the 50 instances available for cycle 1 (top) almost double the 27 available instances for the limonene case (Fig.~\ref{fig:example1}), the predictive power of \textsf{ART} is heavily compromised ($R^2=-0.29$ for cross-validation) by the strong tie of the pathway to host metabolism (fatty acid production), and the scarcity of data. The poor predictions for the test data from cycle 2 (in blue) confirm the lack of predictive power. Adding data from both cycles (1 and 2) improves predictions notably (bottom). These data and model refer to the first pathway in Fig.\ 1B from \citet{Opgenorth2019}. The cases for the other two pathways produce similar conclusions (Figs.~S5 and S6%\ref{FigDodecanol_P2} and~\ref{FigDodecanol_P3}
). $R^2$ and Mean Absolute Error (MAE) values are only for cross-validated mean predictions (black data points). Bars indicate 95\% credible interval of the predictive posterior distribution.}
\label{FigDodecanol_P1}
\end{figure}

The first challenge involved the limited predictive power of machine learning for this case. This limitation is shown by \textsf{ART}'s completely compromised prediction accuracy (Fig.\ \ref{FigDodecanol_P1}). The causes seem to be  {\color{black}twofold}: a small training set and a strong connection of the pathway to the rest of host metabolism. The initial 33 strains (116 instances) were divided into three different designs (Table S4% \ref{TabDodecanol_data}
), decimating the predictive power of \textsf{ART} (Figs.~\ref{FigDodecanol_P1}, S5 %\ref{FigDodecanol_P2} 
and S6%\ref{FigDodecanol_P3}
). Now, it is complicated to estimate the number of strains needed for accurate predictions because that depends on the complexity of the problem to be learnt (see ``Using simulated data to test \textsf{ART}'' section). In this case, the problem is harder to learn than the previous two examples: the mevalonate pathway used in those examples is fully exogenous (i.e.\ built from external genetic parts) to the final yeast host and hence, free of the metabolic regulation that is certainly present for the dodecanol producing pathway. The dodecanol pathway depends on fatty acid biosynthesis which is vital for cell survival (it produces the cell membrane), and has to be therefore tightly regulated\cite{magnuson1993regulation}. This characteristic makes it more difficult to learn its behavior by \textsf{ART} using only dodecanol synthesis pathway protein levels (instead of adding also proteins from other parts of host metabolism).

A second challenge, compounding the first one, involves the inability to reach the target protein levels recommended by \textsf{ART} to increase production. This difficulty precludes not only bioengineering, but also testing the validity of the \textsf{ART} model. For this project, both the mechanistic (RBS calculator\cite{salis2009automated,espah2013translation}) and machine learning-based (EMOPEC\cite{bonde2016predictable}) tools proved to be very inaccurate for bioengineering purposes: e.g.\ a prescribed 6-fold increase in protein expression could only be matched with a 2-fold increase. Moreover, non-target effects (i.e.\ changing the RBS for a gene significantly affects protein expression for other genes in the pathway) were abundant, further adding to the difficulty. While unrelated directly to \textsf{ART} performance, these effects highlight the importance of having enough control over \textsf{ART}'s input (proteins in this case) to obtain satisfactory bioengineering results.  

A third, unexpected, challenge was the inability of constructing several strains in the Build phase due to toxic effects engendered by the proposed protein profiles (Table S4%\ref{TabDodecanol_data}
). This phenomenon materialized through mutations in the final plasmid in the production strain or no colonies after the transformation. The prediction of these effects in the Build phase represents an important target for future ML efforts, in which tools like \textsf{ART} can have an important role. A better understanding of this phenomenon may not only enhance bioengineering but also reveal new fundamental biological knowledge. 

These challenges highlight the importance of carefully considering the full experimental design before leveraging machine learning to guide metabolic engineering.

\subsection{Conclusion}

\textsf{ART} is a tool that not only provides synthetic biologists easy access to machine learning techniques, but can also systematically guide bioengineering and quantify uncertainty. \textsf{ART} takes as input a set of vectors of measurements (e.g.\ a set of proteomics measurements for several proteins, or transcripts for several genes) along with their corresponding systems responses (e.g.\ associated biofuel production) and provides a predictive model, as well as recommendations for the next round (e.g.\ new proteomics targets predicted to improve production in the next round).  

\textsf{ART} combines the methods from the scikit-learn library with a novel Bayesian ensemble approach and MCMC sampling, and is optimized for the conditions encountered in metabolic engineering: small sample sizes, recursive DBTL cycles and the need for uncertainty quantification. \textsf{ART}'s approach involves an ensemble where the weight of each model is considered a random variable with a probability distribution inferred from the available data. Unlike other approaches, this method does not require the ensemble models to be probabilistic in nature, hence allowing us to fully exploit the popular scikit-learn library to increase accuracy by leveraging a diverse set of models. This weighted ensemble model produces a simple, yet powerful, approach to quantify uncertainty (Fig.~\ref{fig:example1}), a critical capability when dealing with small data sets and a crucial component of AI in biological research\cite{begoli2019need}. While \textsf{ART} is adapted to synthetic biology's special needs and characteristics, its implementation is general enough that it is easily applicable to other problems of similar characteristics. \textsf{ART} is perfectly integrated with the Experiment Data Depot \cite{Morrell2017}  and the Inventory of Composable Elements \cite{Ham2012}, forming part of a growing family of tools that standardize and democratize synthetic biology. 

We have showcased the use of \textsf{ART} in a case with synthetic data sets and three real metabolic engineering cases from the published literature. The synthetic data case involves data generated for several production landscapes of increasing complexity and dimensionality. This case allowed us to test \textsf{ART} for different levels of difficulty of the production landscape to be learnt by the algorithms, as well as different numbers of DBTL cycles. We have seen that while easy landscapes provide production increases readily after the first cycle, more complicated ones require $>$5 cycles to start producing satisfactory results (Fig.~\ref{fig:F_E}). In all cases, results improved with the number of DBTL cycles, underlying the importance of designing experiments that continue for $\sim$10 cycles rather than halting the project if results do not improve in the first few cycles.

The demonstration cases using real data involve engineering \emph{E.\ coli} and \emph{S.\ cerevisiae} to produce the renewable biofuel limonene, synthesize metabolites that produce hoppy flavor in beer, and generate dodecanol from fatty acid biosynthesis. Although we were able to produce useful recommendations with as low as 27 (limonene, Fig.~\ref{fig:example1}) or 19 (hopless beer, Fig.~\ref{fig:example2}) instances, we also found situations in which larger amounts of data (50 instances) were insufficient for meaningful predictions (dodecanol, Fig.~\ref{FigDodecanol_P1}). It is impossible to determine a priori how much data will be necessary for accurate predictions, since this depends on the difficulty of the relationships to be learnt (e.g.\ the amount of coupling between the studied pathway and host metabolism). However, one thing is clear---two DBTL cycles (which was as much as was available for all these examples) are rarely sufficient for guaranteed convergence of the learning process. We do find, though, that accurate quantitative predictions are not required to effectively guide bioengineering---our ensemble approach can successfully leverage qualitative agreement between the models in the ensemble to compensate for the lack of accuracy (Fig.~\ref{fig:MLcomp}). Uncertainty quantification is critical to gauge the reliability of the predictions (Fig.~\ref{fig:example1}), anticipate the smoothness of the recommendation process through several DBTL cycles (Figs S3 and S4), and effectively guide the recommendations towards the least understood part of the phase space (exploration case, Fig.~\ref{fig:recommendations}). We have also explored several ways in which the current approach (mapping --omics data to production) can fail when the underlying asssumptions break down. Among the possible pitfalls is the possibility that recommended target protein profiles cannot be accurately reached, since the tools to produce specified protein levels are still imperfect; or because of biophysical, toxicity or regulatory reasons. These areas need further investment in order to accelerate bioengineering and make it more reliable, hence enabling design to a desired specification.  

While \textsf{ART} is a useful tool in guiding bioengineering, it represents just an initial step in applying machine learning to synthetic biology. Future improvements under consideration include adding a pathway cost (\$) function, %enabling categorical/discrete input variables, 
the inclusion of classification problems, adding new optimization methods (e.g.\ to include the case of discrete input variables), incorporating covariance of level-0 models into the ensemble model, and incorporating input space errors into learners.
These may not be the preferred list of improvements for every user, so \textsf{ART}'s dual license allows for modification by third parties for research purposes, as long as the modifications are offered to the original repository. Hence, users are encouraged to enhance it in ways that satisfy their needs. A commercial use license is also available (see below for details).  

\textsf{ART} provides effective decision-making in the context of synthetic biology and facilitates the combination of machine learning and automation that might disrupt synthetic biology\cite{Carbonell2019}. Combining ML with recent developments in macroscale lab automation\cite{granda2018controlling,unthan2015bioprocess}, microfluidics\cite{le2018novel, heinemann2017chip, gardner2013synthetic, iwai2018automated, gach2016droplet} and cloud labs\cite{hayden2014automated} may enable self-driving laboratories\cite{hase2019next,hamedirad2019towards}, which augment automated experimentation platforms with artificial intelligence to facilitate autonomous experimentation. We believe that fully leveraging AI and automation can catalyze a similar step forward in synthetic biology as CRISPR-enabled genetic editing, high-throughput multi-omics phenotyping, and exponentially growing DNA synthesis capabilities have produced in the recent past.

\section*{Competing interests}
  The authors declare that they have no competing interests. 

\section*{Author's contributions}

Z.C., T.R.\ and H.G.M.\ conceived the original idea. T.R.\ and Z.C.\ developed methodology, designed the software, wrote the code and performed computer experiments. T.R.\ designed simulated benchmarks and performed numerical experiments. T.R.\ analyzed all results. K.W.\ wrote tests and documented the code. H.G.M.\ and T.R.\ wrote the paper.

\section{Data availability}
The experimental data analyzed in this paper can be found both in the Experiment Data Depot\cite{Morrell2017}, in the following studies:

% \begin{table}[!h]
\begin{tabular}{ll}
 Study         & Link      \\ \hline
Biofuel & \url{https://public-edd.jbei.org/s/pcap/}     \\
Hopless beer & \url{https://public-edd.agilebiofoundry.org/s/hopless-beer/}  \\
& \url{https://public-edd.agilebiofoundry.org/s/hopless-beer-cycle-2/}  \\
Dodecanol & \url{https://public-edd.jbei.org/s/ajinomoto/}    
\end{tabular}
% \caption{}
% \label{}
% \end{table}

\medskip
\noindent and as .csv files in the Github repository (see below).

\section{Software availability}

\textsf{ART}'s dual license allows for free non-commercial use for academic institutions. Modifications should be fed back to the original repository for the benefit of all users. A separate commercial use license is available from Berkeley Lab (ipo@lbl.gov). See \url{https://github.com/JBEI/ART} for software and licensing details.

%{\color{blue} REVIEWERS may access the source code at the private repository: 

%\url{https://github.com/JBEI/AutomatedRecommendationTool} 

%with usernames GuestReviewer1 and GuestReviewer2, and passwords grev1artG and grev2artG, respectively.}

%%%%%%%%%%%%%%%%%%%%%%%%%%%%%%%%%%%%%%%%%%%%%%%%%%%%%%%%%%%%%%%%%%%%%
%% The "Acknowledgement" section can be given in all manuscript
%% classes.  This should be given within the "acknowledgement"
%% environment, which will make the correct section or running title.
%%%%%%%%%%%%%%%%%%%%%%%%%%%%%%%%%%%%%%%%%%%%%%%%%%%%%%%%%%%%%%%%%%%%%
\begin{acknowledgement}
This work was part of the DOE Joint BioEnergy Institute (\url{http:// www.jbei.org}) and part of the Agile BioFoundry (\url{http://agilebiofoundry.org}) supported by the U.\ S.\ Department of Energy, Energy Efficiency and Renewable Energy, Bioenergy Technologies Office, {\color{black}Office of Science,} through contract DE-AC02-05CH11231 between Lawrence Berkeley National Laboratory and the U.\ S.\ Department of Energy. The United States Government retains and the publisher, by accepting the article for publication, acknowledges that the United States Government retains a nonexclusive, paid-up, irrevocable, worldwide license to publish or reproduce the published form of this manuscript, or allow others to do so, for United States Government purposes. The Department of Energy will provide public access to these results of federally sponsored research in accordance with the DOE Public Access Plan (\url{http://energy.gov/downloads/doe-public-access-plan}).
This research is also supported by the Basque Government through the BERC 2014-2017 program and by Spanish Ministry of Economy and Competitiveness MINECO: BCAM Severo Ochoa excellence accreditation SEV-2013-0323.

We acknowledge and thank Peter St.\ John, Joonhoon Kim, Amin Zargar, Henrique De Paoli, Sean Peisert and Nathan Hillson for reviewing the manuscript and for their helpful suggestions.

\end{acknowledgement}

%%%%%%%%%%%%%%%%%%%%%%%%%%%%%%%%%%%%%%%%%%%%%%%%%%%%%%%%%%%%%%%%%%%%%
%% The appropriate \bibliography command should be placed here.
%% Notice that the class file automatically sets \bibliographystyle
%% and also names the section correctly.
%%%%%%%%%%%%%%%%%%%%%%%%%%%%%%%%%%%%%%%%%%%%%%%%%%%%%%%%%%%%%%%%%%%%%

\clearpage

\section{Supplementary material}
\beginsupplement

\subsection{Mathematical Methods}

\subsubsection{Markov Chain Monte Carlo sampling}\label{SupSec:MCMC}
The posterior distribution $p(\boldsymbol\theta|\mathcal D)$ (probability that the parameters $\boldsymbol\theta$ fit the data $\mathcal D$, used in Eq.~\ref{Eq:post_pred}) 
is obtained by applying Bayes' formula, i.e.\ it is defined through a prior $p(\boldsymbol\theta)$ and a likelihood function $p(\mathcal D|\boldsymbol\theta)$ as
$$p(\boldsymbol\theta|\mathcal D)\propto p(\mathcal D|\boldsymbol\theta)p(\boldsymbol\theta).$$
We define the prior to be $p(\boldsymbol\theta)=p(\boldsymbol w)p(\sigma)$, where $p(\boldsymbol w)$ is a Dirichlet distribution with uniform parameters, which ensures the constraint on weights (i.e. they all add to one) is satisfied, and $p(\sigma)$ is a half normal distribution with mean and standard deviation set to 0 and 10, respectively. The likelihood function follows directly from Eq.~\eqref{Eq:response_var}
as
$$p(\mathcal D|\boldsymbol\theta)=\prod_{n=1}^Np(y_n|\mathbf x_n,\boldsymbol\theta), \; \; \; \; p(y_n|\mathbf x_n,\boldsymbol\theta)= \frac{1}{\sigma\sqrt{2\pi}}\exp\left\{-\frac{(y_n-\boldsymbol w^T\mathbf f(\mathbf x_n))^2}{2\sigma^2}\right\}.$$

\subsubsection{Expected value and variance for ensemble model}

From Eq.~\eqref{Eq:response_var}
, we can easily compute the expected value
\begin{equation}\label{Eq:pp_mean}
\mathbb E(y)=\mathbb E(\boldsymbol{w}^T \mathbf f+\varepsilon)=\mathbb E(\boldsymbol{w})^T{\mathbf f}
\end{equation}
and variance 
%$$\mathrm{Var}(y)={\mathbf F}^T\mathrm{Var}(\boldsymbol{\beta}_0){\mathbf F}+ \mathbb E(\boldsymbol{\beta})^T\Sigma \mathbb E(\boldsymbol{\beta})+\mathbb 1^T(\mathrm{Var}(\boldsymbol{\beta})\odot \Sigma)\mathbb 1 +\mathrm{Var}(\varepsilon).$$
\begin{equation}\label{Eq:pp_var}
\mathrm{Var}(y)={\mathbf f}^T\mathrm{Var}(\boldsymbol{w}){\mathbf f} +\mathrm{Var}(\varepsilon)
\end{equation}
%$\boldsymbol{\beta}_0=[\alpha \; \boldsymbol{\beta}]$. 
of the response, which will be needed for are used in the optimization phase in order to create the surrogate function $G(\bold x)$ (Eq.~\ref{Eq:G_function}). The expected value and variance of $\boldsymbol{w}$ and $\varepsilon$ are estimated through sample mean and variance using samples from the posterior distribution $p(\boldsymbol\theta|\mathcal D)$.

Please note that although we have modeled $p(y| \mathbf x^*,\boldsymbol\theta)$ to be Gaussian (Eq.~\ref{Eq:response_var}), the predictive posterior distribution $p(y| \mathbf x^*,\mathcal D)$ (Eq.~\ref{Eq:post_pred}) is not Gaussian due to the complexity of $p(\boldsymbol\theta|\mathcal D)$ arising from the data and other constraints.

It is important to note that our modeling approach provides quantification of both epistemic and aleatoric uncertainty, through the first and second terms in Eq.~\eqref{Eq:pp_var}, respectively.
Epistemic (systematic) uncertainty accounts for uncertainty in the model and aleatoric (statistical) uncertainty describes the amount of noise inherent in the data \cite{Kendall2017}.
While epistemic uncertainty can be eventually explained away given enough data, we need to model it accurately in order to properly capture situations not encountered in the training set. Modeling epistemic uncertainty is therefore important for small data problems, while aleatoric uncertainty is more relevant for large data problems. In general, it is useful to characterize the uncertainties within a model, as this enables understanding which uncertainties have the potential of being reduced \cite{Kiureghian2009}.

\subsubsection{Related work and novelty of our ensemble approach}

Our ensemble approach is based on stacking\cite{Breiman1996}---a method where different ensemble members are trained on the same training set and whose outputs are then combined, as opposed to techniques that manipulate the training set (e.g.\ bagging \cite{Breiman1996b}) or those that sequentially add new models into the ensemble (e.g.\ boosting\cite{Freund1997}).
Different approaches for constructing ensemble of models using the Bayesian framework have been already considered.
%%%%%%%%%%%%%%%%%%%%%%%%%%%%%%%%%%%%%%%%%%%
For example, Bayesian Model Averaging (BMA) \cite{Hoeting1999} builds an ensemble model as a linear combination of the individual members in which the weights are given by the posterior probabilities of models. The weights therefore crucially depend on marginal likelihood under each model, which is challenging to compute. BMA accounts for uncertainty about which model is correct but assumes that only one of them is, and as a consequence, it has the tendency of selecting the one model that is the closest to the generating distribution.
% BMA has the tendency of converge toward giving all of the weight to a single model
%will asymptotically select the single model from the available ones. 
%BMA ensemble effectively shrinks to contain only a single model
%%%%%%%%%%%%%%%%%%%%%%%%%%%%%%%%%%%%%%%%%%%
Agnostic Bayesian learning of ensembles \cite{Lacoste2014} differs from BMA in the way the weights are calculated. Instead of finding the best predictor from the model class (assuming that the observed data is generated by one of them), this method aims to find the best predictor in terms of the lowest expected loss. The weights are calculated as posterior probability that each model is the one with the lowest loss.
%%%%%%%%%%%%%%%%%%%%%%%%%%%%%%%%%%%%%%%%%%%
Bayesian model combination (BMC)\cite{Monteith2011} seeks the combination of models that is closest to the generating distribution by heavily weighting the most probable combination of models, instead of doing so for the most probable one.
BMC samples from the space of possible ensembles by randomly drawing weights from a Dirichlet distribution with uniform parameters. 
%%%%%%%%%%%%%%%%%%%%%%%%%%%%%%%%%%%%%%%%%%%
The Bayesian Additive Regression Trees (BART)\cite{Chipman2006} method is one of the homogeneous ensemble approaches. It models the ensemble as a (nonweighted) sum of regression trees whose parameters, and ensemble error standard deviation, are defined thought their posterior distributions given data and sampled using MCMC.
%%%%%%%%%%%%%%%%%%%%%%%%%%%%%%%%%%%%%%%%%%%
\citet{Yao2018} suggest a predictive model in terms of a weighted combination of predictive distributions for each probabilistic model in the ensemble. This approach can be seen as a generalization of stacking for point estimation to predictive distributions.
%%%%%%%%%%%%%%%%%%%%%%%%%%%%%%%%%%%%%%%%%%%

All of these models, except of BMC and our model, have weights being point estimates, obtained usually by minimizing some error function. In contrast, we  define them as random variables, and in contrast to BMC, our weights are defined through full joint posterior distribution given data. BMC is the closest in design to our approach, but it was formulated only in the context of classifiers. Only BART does include a random error term in the ensemble, apart from our model.
Unlike BMA, BMC or models of \citet{Yao2018}, \citet{Chipman2006}, our approach does not require that the predictors are themselves probabilistic, and therefore can readily leverage various scikit-learn models. The main differences are summarized in Table \ref{Tab:method_differences}.

\begin{table}[!h]
	\centering
	\begin{tabular}{lcccccc}
		\multirow{2}{*}{Method}  & Weighted  & Probabilistic  & Probabilistic & \multirow{2}{*}{Regression} & \multirow{2}{*}{Classification} & Ensemble \\
		&  average &  base models & weights &  &  & error term \\\hline
		BMA\cite{Hoeting1999} & \ding{52} &\ding{52}  & \ding{56}& \ding{52}& \ding{52}&\ding{56}\\
		BMC\cite{Monteith2011} & \ding{52}&\ding{52}& \ding{52}\ding{56}& \ding{56}& \ding{52}&\ding{56}\\
		BART\cite{Chipman2006} &  \ding{56}&\ding{52}& \ding{56}& \ding{52}& \ding{56}&\ding{52}\\
		Stacking predictive  & \multirow{2}{*}{\ding{52}}&\multirow{2}{*}{\ding{52}}& \multirow{2}{*}{\ding{56}}& \multirow{2}{*}{\ding{52}} & \multirow{2}{*}{\ding{52}}&\multirow{2}{*}{\ding{56}}\\
		distributions\cite{Yao2018} &  & &&& &\\
		Agnostic Bayes\cite{Lacoste2014} & \ding{52} &\ding{52}\ding{56} & \ding{56}&\ding{52}& \ding{52}&\ding{56}\\
		{\color{black}ART (this work)} & \ding{52}&\ding{56} &\ding{52}&\ding{52}& \ding{56}& \ding{52}\\
	\end{tabular}
	\caption{Feature differences between Bayesian based ensemble modeling approaches.}
	\label{Tab:method_differences}
\end{table}

Although our model has some features that {\color{black}have been} previously considered in the literature, {\color{black}the approach presented here is, to the best of our knowledge, novel in the fact} that the metalearner is modeled as a Bayesian linear regression model, whose parameters are inferred from data combined with a prior that satisfies the constraints on the `voting' nature of ensemble learners.

\subsubsection{Input space set $\mathcal{B}$}
The bounds for the input space $\mathcal{B}$ for $G(\bold x)$ (Eq.~\ref{Eq:G_function}) can be provided by the user (see details in the %\ref{Sec:Implementation}
Implementation section, Table~\ref{Tab:ART_Input}). Otherwise, default values are computed from the input data defining the feasible space as:
\begin{equation}
\begin{split}
\mathcal B=\{\mathbf{\tilde{x}}\in \mathbb R^D| \;\; L_d - \Delta_d \le \tilde{x}^d \le U_d + \Delta_d\;\;,\; d=1,\dots,D\} \\
\Delta_d = (U_d-L_d)\epsilon \;\; ; \;\; U_d = \mathop{\textrm{max}}_{1\le n \le N} (x_n^d) \;\; ; \;\; L_d = \mathop{\textrm{min}}_{1\le n \le N} (x_n^d)\\
(\mathbf{x}_n,y_n) \in \mathcal{D},\; n=1,\dots,N  
\end{split}
\end{equation}
The restriction of input variables to the set $\mathcal B$ reflects our assumption that the predictive models performs accurately enough only on an interval that is enlarged by a factor $\epsilon$ around the minimum and maximum values in the data set ($\epsilon= 0.05$ in all calculations).

\subsubsection{Success probability calculation} 

Our probabilistic model enables us to estimate the probability of success for the provided recommendations. Of practical interest are the probability that a single recommendation is successful and the probability that at least one recommendation of several provided is successful.

Success is defined differently for each of the three cases considered in Eq.~\eqref{Eq:G_function}
: maximization, minimization and specification. For maximization, success involves obtaining a response $y$ higher than the success value $y^*$ defined by the user (e.g.\ the best production so far improved by a factor of 20\%). For minimization, success involves obtaining a response lower than the success value $y^*$. For the specification case, it involves obtaining a response that is as close as possible to the success value $y^*$. 

Formally, we define success for response $y$ through the set $\mathcal S=\{y| y \sim p_{\mathcal S}(y)\}$, where the probability distribution for success is
\begin{equation} \label{Eq:succ_prob_general}
p_{\mathcal S}(y) = \left\{ \begin{array}{lc} 
\mathcal U(y^*, U) & \text{(maximization case)} \\
\mathcal U(L, y^*) & \text{(minimization case)} \\
\mathcal N(y^*, \sigma^2_{y^*}) & \text{(specification case)},
\end{array} \right.
\end{equation}
where $\mathcal U$ is the uniform distribution ($\mathcal U(a,b) = 1/(b-a)$ if $a<y<b$; $0$ otherwise), $L$ and $U$ are its corresponding lower and upper bounds, and $\sigma^2_{y^*}$ is the variance of the normal distribution $\mathcal N$ around the target value $y^*$ for the specification case.

The probability that a recommendation succeeds is given by integrating the probability that input $\mathbf x^r$ gives a response $y$ (i.e.\ full predictive posterior distribution from Eq.~\ref{Eq:post_pred}), times the probability that response $y$ is a success
$$p(\mathcal S|\mathbf x^r)=\int p_{\mathcal S}(y) p(y|\mathbf x^r,\mathcal D)\mathrm d y.$$
This success probability is approximated using draws from the posterior predictive distribution as
\begin{equation} \label{Eq:succ_prob_approx}
p(\mathcal S|\mathbf x^r) \approx \left\{ \begin{array}{lc} 
\frac{1}{N_s}\sum_{i=1}^{N_s}\mathbb{I}_{\mathcal S}(y_i)  & \text{(maximization/minimization case)} \\
\frac{1}{N_s}\sum_{i=1}^{N_s}\mathcal{N}(y_i;y^*,\sigma_{y^*}^2) & \text{(specification case)}
\end{array} \right.
\end{equation}
where $y_i\sim p(y|\mathbf x^r,\mathcal D), i=1,\dots,N_s,$ and
$\mathbb{I}_{\mathcal A}(y)= 1$ if  $y \in {\mathcal A}$, $0$ if  $y \not \in {\mathcal A}$.
% for now we can provide only success probability for individual recommendations

In case of multiple recommendations $\{\mathbf x^r\}\equiv \{\mathbf x^r\}_{r=1}^{N_r}$, we provide the probability of success for at least one of the recommendations only for maximization and minimization types of objectives. This probability is calculated as one minus the probability $p(\mathcal F|\{\mathbf x^r\})$ that all recommendations fail, where
$$p(\mathcal F|\{\mathbf x^r\})\approx \frac{1}{N_s}\sum_{i=1}^{N_s}\mathbb{I}_{\mathcal F}(\{y_i^r\}), \; \{y_i^r\}\sim p(y|\{\mathbf x^r\},\mathcal D), i=1,\dots,N_s, r=1,\dots,N_r,$$
and the \emph{failure} set $\mathcal F=\left\{\{y^r\} |y^r \notin \mathcal S, \forall r=1,\dots,N_r \right\}$ consists of outcomes that are not successes for all of the recommendations. Since the chosen recommendations are not necessarily independent, we sample $\{y_i^r\}$ jointly for all $\{\mathbf x^r\}$, i.e.\ $i$-th sample has the same model parameters ($w_i,\sigma_i, \varepsilon_{ij}\sim\mathcal N(0,\sigma_i^2)$ from Eq.\ \ref{Eq:response_var}) for all recommendations.

\subsubsection{Multiple response variables} 

For multiple response variable problems (e.g.\ trying to hit a predetermined value of metabolite \emph{a} and metabolite \emph{b} simultaneously, as in the case of the hopless beer), we assume that the response variables are conditionally independent given input vector $\mathbf x$, and build a separate predictive model $p_j(y_j|\mathbf x,\mathcal D)$ for each variable $y_j, j=1,\dots,J$. We then define the objective function for the optimization phase as
$$G(\mathbf x)=(1-\alpha)\sum_{j=1}^J\mathbb E(y_j)+\alpha \sum_{j=1}^J\mathrm{Var}(y_j)^{1/2}$$
in case of maximization, and analogously adding the summation of expectation and variance terms in the corresponding functions for minimization and specification objectives (Eq.\ \ref{Eq:G_function}). The probability of success for multiple variables is then defined as 
$$p(\mathcal S_1,\dots,\mathcal S_J|\mathbf x)=\prod_{j=1}^J p(\mathcal S_j|\mathbf x^r)$$

Future work will address the limitation of the independence assumption and take into account possible correlations among multiple response variables.

\subsection{Implementation}\label{SuplSec:Implementation}

\begin{figure}[h]
	\begin{center}
		\includegraphics[width=1.\columnwidth]{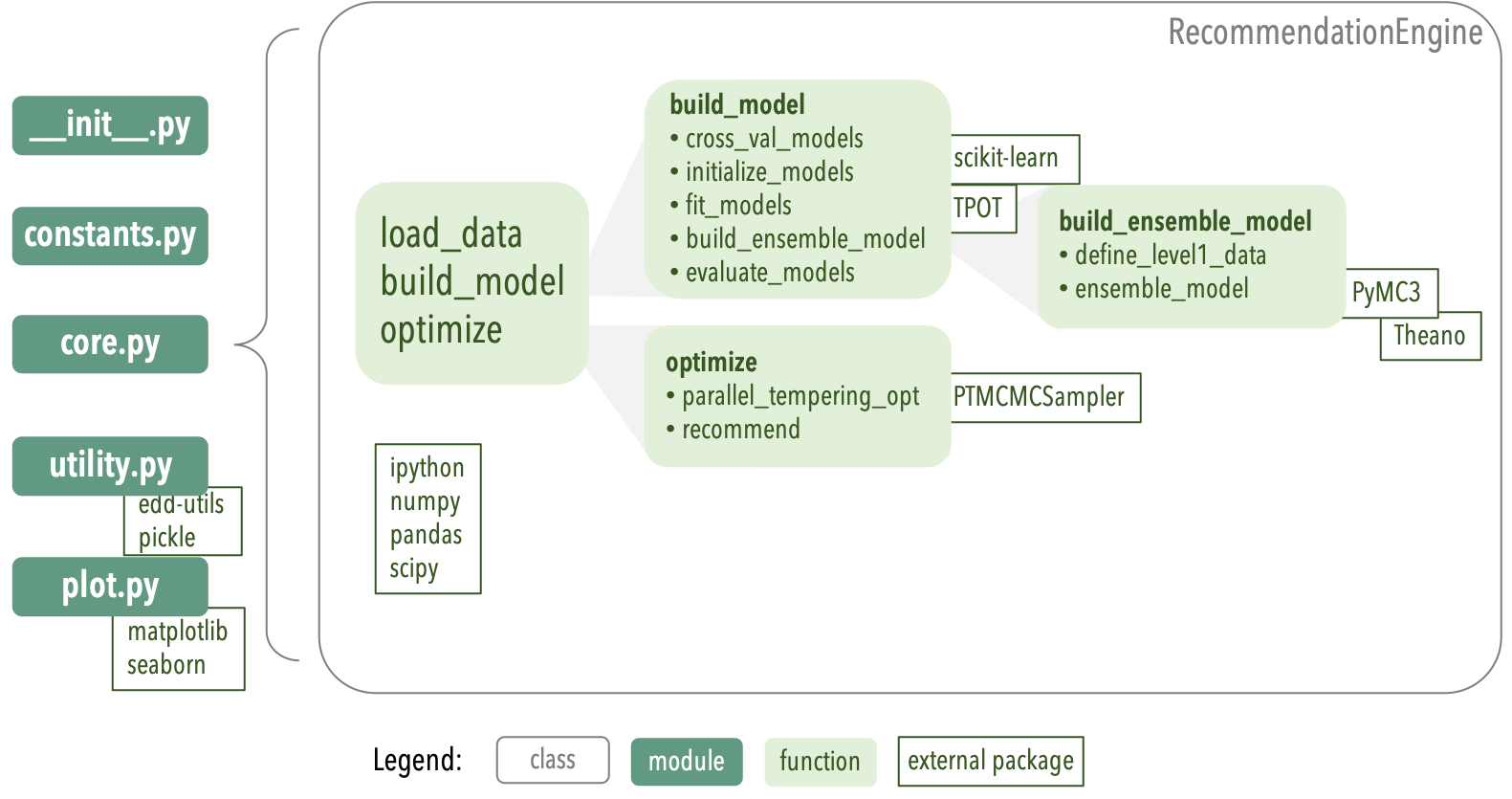}
	\end{center}
	\caption{The main \textsf{ART} source code structure and dependencies.}
	\label{fig:codeStructure}
\end{figure}

\subsubsection{Modules}
\texttt{core.py} is the core module that defines the class \texttt{RecommendationEngine} with functions for loading data (into the format required for machine learning models), building predictive models and optimization (Fig.~\ref{fig:codeStructure}). 

The module \texttt{constants.py} contains assignments to all constants appearing throughout all other modules. Those include default values for some of the optional user input parameters (Table \ref{Tab:ART_Input}), hyperparameters for \texttt{scikit-learn} models and simulation setups for \texttt{PyMC3} and \texttt{PTMCMCSampler} functions.

Module \texttt{utilities.py} is a suite of functions that facilitate \textsf{ART}'s computations but can be used independently. It includes functions for loading studies (using EDD through \texttt{edd-utils} or directly from files), metrics for evaluation of predictive models, identifying and filtering noisy data, etc.

Module \texttt{plot.py} contains a set of functions for visualization of different quantities obtained during an \textsf{ART} run, including functions of relevance to final users (e.g.\ true vs.\ predicted values) as well as those providing insights into intermediate steps (e.g.\ predictive models surfaces, space exploration from optimization, recommendations distributions).

All modules can be easily further extended by future contributors to \textsf{ART}.

\subsubsection{Importing a study}

Studies can be loaded directly from EDD by calling a function from the \texttt{utility.py} module that relies on \texttt{edd-utils} package:
%\begin{minted}
%[bgcolor=bg,fontsize=\footnotesize]{python}
%dataframe = load_study(edd_study_slug=edd_study_slug,edd_server=edd_server)
%\end{minted}

\noindent{\small \texttt{dataframe = load\_study(edd\_study\_slug=edd\_study\_slug,edd\_server=edd\_server)}
}

\noindent The user should provide the study slug (last part of the study web address) and the url to the EDD server.  
Alternatively, a study can be loaded from an EDD-style .csv file, by providing a path to the file and calling the same function:
%\begin{minted}[bgcolor=bg,fontsize=\footnotesize]{python}
%dataframe = load_study(data_file=data_file)
%\end{minted}

\noindent{\small \texttt{dataframe = load\_study(data\_file=data\_file})
}

\noindent The .csv file should have the same format as an export file from EDD, i.e.\ it should contain at least \texttt{Line Name} column, \texttt{Measurement Type} column with names of input and response variables, and \texttt{Value} column with numerical values for all variables.

\noindent Either approach will return a pandas dataframe containing all information in the study, which can be pre-processed before running \textsf{ART}, if needed.

\subsubsection{Running \textsf{ART}}

\textsf{ART} can be run by instantiating an object from the \texttt{RecommendationEngine} class by: 
%\begin{minted}[bgcolor=bg,fontsize=\footnotesize]{python}
%art = RecommendationEngine(dataframe, **art_params)
%\end{minted}

\noindent{\small \texttt{art = RecommendationEngine(dataframe, **art\_params)}
}

The first argument is the dataframe created in the previous step (from an EDD study or data file import). If there is no data preprocessing, the dataframe is ready to be passed as an argument. Otherwise, the user should make sure that the dataframe contains at least the required columns: \texttt{Line Name}, \texttt{Measurement Type} and \texttt{Value}. Furthermore, line names should always contain a hyphen (``\texttt{-}'') denoting replicates (see Table \ref{Tab:Valid_line_names}), and this character should be exclusively used for this purpose (this point is critical for creating partitions for cross-validation).

\begin{table}[!h]
	\centering
	\begin{tabular}{ll}
		\ding{52} Valid & \ding{56} Non valid \\ \hline
		\texttt{LineNameX-1} & \texttt{LineNameX1}\\
		\texttt{LineNameX-2}& \texttt{LineNameX2}\\
		\texttt{LineNameX-r1}&  \texttt{Line-NameX1}\\
		\texttt{LineNameX-r2}& \texttt{Line-NameX2}\\
		\texttt{LineNameX-R1}& \texttt{Line-Name-X1}\\
		\texttt{LineNameX-R2} & \texttt{Line-Name-X2}\\
		\dots & \dots
	\end{tabular}
	\caption{Valid and non valid examples of entries of the \texttt{Line Name} column in the dataframe passed to start an \textsf{ART} run.}
	\label{Tab:Valid_line_names}
\end{table}

The second argument is a dictionary of key-value pairs defining several required and optional keyword arguments (summarized in Table \ref{Tab:ART_Input}) for generation of an \texttt{art} object.

\begin{table}[!h]
	\centering
	\begin{tabular}{ll}
		Name & Meaning \\ \hline
		\texttt{input\_var} & List of input variables*\\
		\texttt{bounds\_file} & Path to the file with upper and lower bounds for each input variable \\
		& (default \texttt{None}) \\
		\texttt{response\_var} & List of response variables*\\
		\texttt{build\_model} & Flag for building a predictive model (default \texttt{True}) \\
		\texttt{cross\_val} & Flag for performing cross-validation (default \texttt{False}) \\
		\texttt{ensemble\_model} & Type of the ensemble model (default \texttt{`BW'}) \\
		\texttt{num\_models} & Number of level-0 models (default 8) \\
		\texttt{recommend} & Flag for performing optimization and providing recommendations \\
		& (default \texttt{True}) \\
		\texttt{objective} & Type of the objective (default \texttt{`maximize'}) \\
		\texttt{threshold} & Relative threshold for defining success (default 0) \\
		\texttt{target\_value} & Target value for the specification objective (default \texttt{None})\\
		\texttt{num\_recommendations} & Number of recommendations for the next cycle (default 16)\\
		\texttt{rel\_eng\_accuracy} & Relative engineering accuracy or required relative distance between \\
		& recommendations (default 0.2)\\
		\texttt{niter} & Number of iterations to use for $T=1$ chain in parallel tempering \\
		& (default 100000)\\ 
		\texttt{alpha} & Parameter determining the level of exploration during the \\  
		& optimization (value between 0 and 1, default \texttt{None} corresponding to 0)\\ 
		\texttt{output\_directory} & Path of the output directory with results (default \\
		& \texttt{../results/{response\_var}\_{time\_suffix}})\\
		\texttt{verbose} & Amount of information displayed (default 0)\\ 
		\texttt{seed} & Random seed for reproducible runs (default \texttt{None})\\
	\end{tabular}
	\caption{\textsf{ART} input parameters. Required parameters are marked with an asterisk.}
	\label{Tab:ART_Input}
\end{table}

\medskip
\emph{Building the model}

The level-0 models are first initialized and then fitted through the \texttt{\_initialize\_models} and \texttt{\_fit\_models} functions respectively, which rely on the \texttt{scikit-learn} and \texttt{tpot} packages. 
% \textcolor{blue}{Level-0 models are initialized once, using the \texttt{scikit-learn} and \texttt{tpot} libraries, and hashed to the \texttt{\_art\_models} structure to avoid potentially expensive repeat initialization of unused models. Relevant level-0 models are then retrieved, initialized, and fitted with the \texttt{\_initialize\_models} and \texttt{\_fit\_models} functions respectively.} 
To build the final predictive model, first the level-1 data is created by storing cross-validated predictions of level-0 models into a \texttt{theano} variable that is shared across the functions from the \texttt{PyMC3} package. Finally, the parameters of the ensemble model are sampled within the function \texttt{\_ensemble\_model}, which stores the inferred model and traces that are later used for predictive posterior probability calculation, as well as first and second moments from the traces, used for estimation of the first two moments of the predictive posterior distribution using Eq.\ \eqref{Eq:pp_mean}--\eqref{Eq:pp_var}.

By default, \textsf{ART} builds the models using all available data and evaluates the final, ensemble model, as well as all level-0 models, on the same data. Optionally, if specified by the user through the input flag \texttt{cross\_val}, \textsf{ART} will evaluate the models on 10-fold cross-validated predictions, through the function \texttt{\_cross\_val\_models}. This computation lasts roughly 10 times longer. Evaluating models on new data, unseen by the models, can also be done by calling: 
%\begin{minted}[bgcolor=bg,fontsize=\footnotesize]{python}
%art.evaluate_models(X=X_new, y=y_new)
%\end{minted}
%{\color{blue}Add which performance metrics are reported (MAE, MSE, $R^2$...).}

\noindent{\small \texttt{art.evaluate\_models(X=X\_new, y=y\_new)}
}

\medskip
\emph{Optimization}

\textsf{ART} performs optimization by first creating a set of draws from
%\begin{minted}[bgcolor=bg,fontsize=\footnotesize]{python}
%draws = art.parallel_tempering_opt()
%\end{minted}

\noindent{\small \texttt{draws = art.parallel\_tempering\_opt()}
}

\noindent which relies on the \texttt{PTMCMCSampler} package. Here, an object from the class \texttt{TargetModel} is created. This class provides a template for and can be replaced by other types of objective functions (or target distributions) for parallel tempering type of optimization, as long as it contains functions defining loglikelihood and logprior calculation (see Eq.~\ref{Eq:target_distr}). Also, the whole optimization procedure may well be replaced by an alternative routine. For example, if the dimension of the input space is relatively small, a grid search could be performed, or even evaluation of the objective at each point for discrete variables.
Lastly, out of all draws collected by optimizing the specified objective, \textsf{ART} finds a set of recommendations by
%\begin{minted}[bgcolor=bg,fontsize=\footnotesize]{python}
%art.recommend(draws, rel_eng_accuracy=rel_eng_accuracy, distance_type='at_least_one')
%\end{minted}

\noindent{\small \texttt{art.recommend(draws)}
}

\noindent which ensures that each recommendation is different from all others and all input data by a factor of $\gamma$ (\texttt{rel\_eng\_accuracy}) in at least one of the components (see Algorithm \ref{Alg:Recommendations}).

% \subsection{Pseudo algorithm for recommendations}

\begin{algorithm}
	\caption{Choosing recommendations from a set of samples from the target distribution $\pi(\mathbf x)$}
	\label{Alg:Recommendations}
	\begin{algorithmic}[1]
		\State \textbf{Input:} $N_r$: number of recommendations
		\newline
		{\color{white}\textbf{Input:}} $\{\mathbf x_n\}_{n=1}^{N_s}$: samples from $\pi(\mathbf x)$ (Eq.~\ref{Eq:target_distr})
		\newline {\color{white}\textbf{Input:}} $\gamma$: required relative distance for recommendations (relative engineering accuracy)
		\newline {\color{white}\textbf{Input:}} $\mathcal D_{\mathbf x}$: input variable experimental data
		% 		\newline {\color{white}\textbf{Input:}} type of the distance between recommendations
		\State \textbf{Output:} $rec=\{\mathbf x^r\}_{r=1}^{N_r}$: set of recommendations
		\State  ${draws} \leftarrow \{\mathbf x_n\}_{n=1}^{N_s}$ \Comment{remaining draws}
		\State ${rec}=\emptyset$
		\WHILE {$r=1,\dots, N_r$}
		\IF {${draws}=\emptyset$}
		\State $\gamma=0.8\gamma$ and repeat the procedure
		\ELSE
		\State ${\mathbf x^r} \leftarrow$ a sample from ${draws}$ with maximal $G(\mathbf x)$ \Comment{$G(\mathbf x)$ is already calculated}
		\IF {there exists $d\in\{1,\dots,D\}$ s.t. $|x^r_d-x_d|>\gamma$ for all $\mathbf x \in rec\cup \mathcal D_{\mathbf x}$}
		\State  ${rec}=\{rec, \mathbf x^r\}$
		\ENDIF
		\ENDIF
		\State ${draws} \leftarrow draws \; \setminus \{\mathbf x_n \in draws|G(\mathbf x_n)= G(\mathbf x^r)\}$
		\ENDWHILE
		\State \textbf{return} ${rec}$
	\end{algorithmic}
\end{algorithm}

\subsubsection{Documentation and testing}

The \textsf{ART} source code thoroughly conforms to the syntactic and stylistic specifications outlined in the PEP 8 style guide \cite{pep8}. 
Documentation for class methods and utility functions is embedded in docstrings. A brief synopsis of functionality, an optional deeper dive into behavior and use cases, as well as a full description of parameter and return value typing is included in the flexible reStructuredText markup syntax. The documented source code is used in conjunction with the \texttt{Sphinx} package to dynamically generate an API reference.% that is hosted at \url{art.jbei.org} {\color{red} (is this working???)}, along with other information describing the testing and basic use of \textsf{ART}.

The code can be built as a local package for testing and other utility, the dependencies and procedure for which are handled by the \texttt{setup.py} file in accordance with the Setuptools standard for module installation.

A suite of unit and integration tests were written using the \texttt{pytest} library, and are included with the source code under the \texttt{tests} directory. The unit testing was designed to target and rigorously validate individual functions in their handling and output of types. Because of the expensive calls to libraries such as \texttt{scikit-learn} and \texttt{tpot} laden throughout \textsf{ART}'s codebase for model training and validation, unit-tested functions were parameterized with the Boolean flag \texttt{testing} to replace such calls with dummy data that mimic the responses from library code. The resulting suite of unit tests therefore runs rapidly, quickly evaluating the integrity of type handling and output that ensure the smooth hand-off of values between functions in \textsf{ART}. 

Integration tests are likewise implemented with \texttt{pytest}, but rather test a more complete spectrum of expected behavior without silencing calls to library code. Full model training, cross-validation, and evaluation is completed under this test suite, the output of which is validated against known data sets within some reasonable margin to account for stochastic processes.

The instructions to locally run the testing suite can be found in the %hosted
documentation.

%{\color{blue}Mention that input settings, models, variables, recommendations etc are saved either through cvs or pkl files. This helps reproducibility and traceability.}

\subsection{Supplementary figures and tables}

\begin{figure}[!h]
	\begin{center}
		\includegraphics[width=0.9\columnwidth]{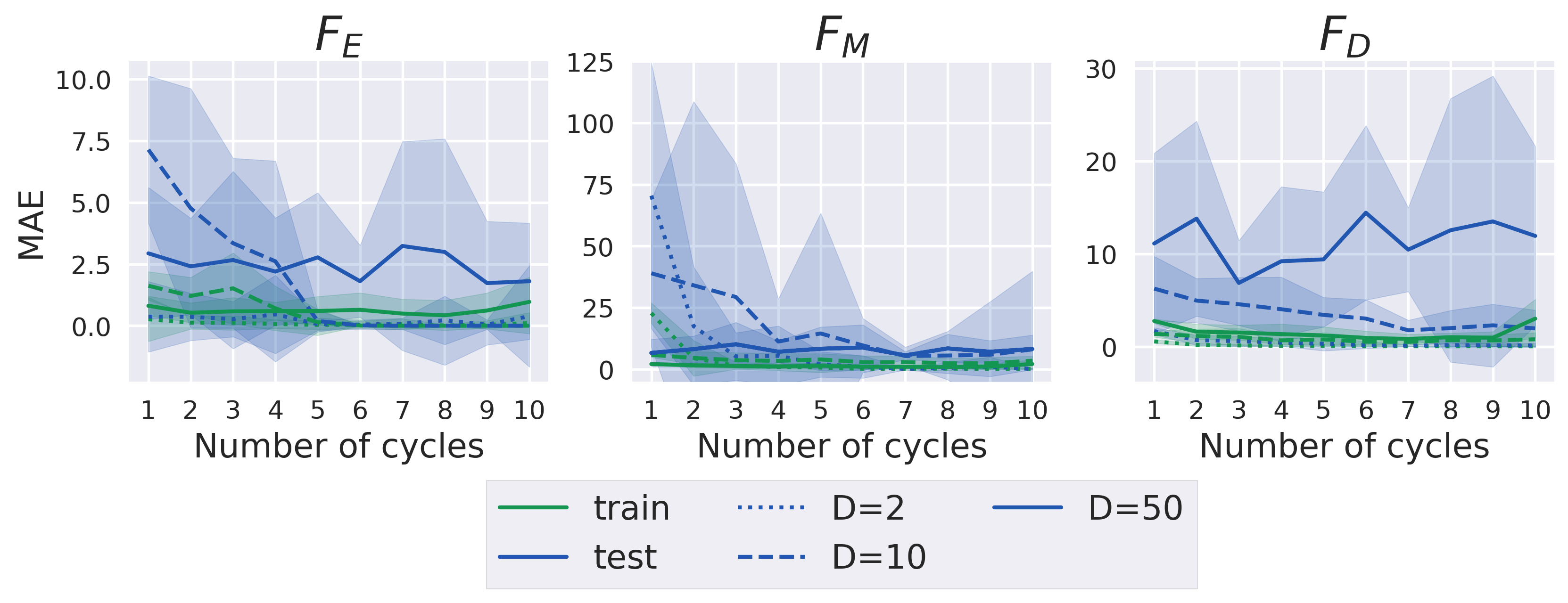}
	\end{center}
	\caption{\textbf{Mean Absolute Error (MAE) for the synthetic data set in Fig.~\ref{fig:F_E}}. 
		Synthetic data is obtained from functions of different levels of complexity (see Table \ref{Tab:functions}), different phase space dimensions (2, 10 and 50), and different amounts of training data (DBTL cycles). The training set involves all the strains from previous DBTL cycles. The test set involves the recommendations from the current cycle. MAE are obtained by averaging the absolute difference between predicted and actual production levels for these strains. MAE decreases significantly as more data (DBTL cycles) are added, with the exception of the high dimension case. In each plot, lines and shaded areas represent the estimated mean values and 95\% confidence intervals, respectively, over 10 repeated runs.}
	\label{FigMAE}
\end{figure}

\begin{figure}[!h]
	\begin{center}
		\includegraphics[width=0.7\columnwidth]{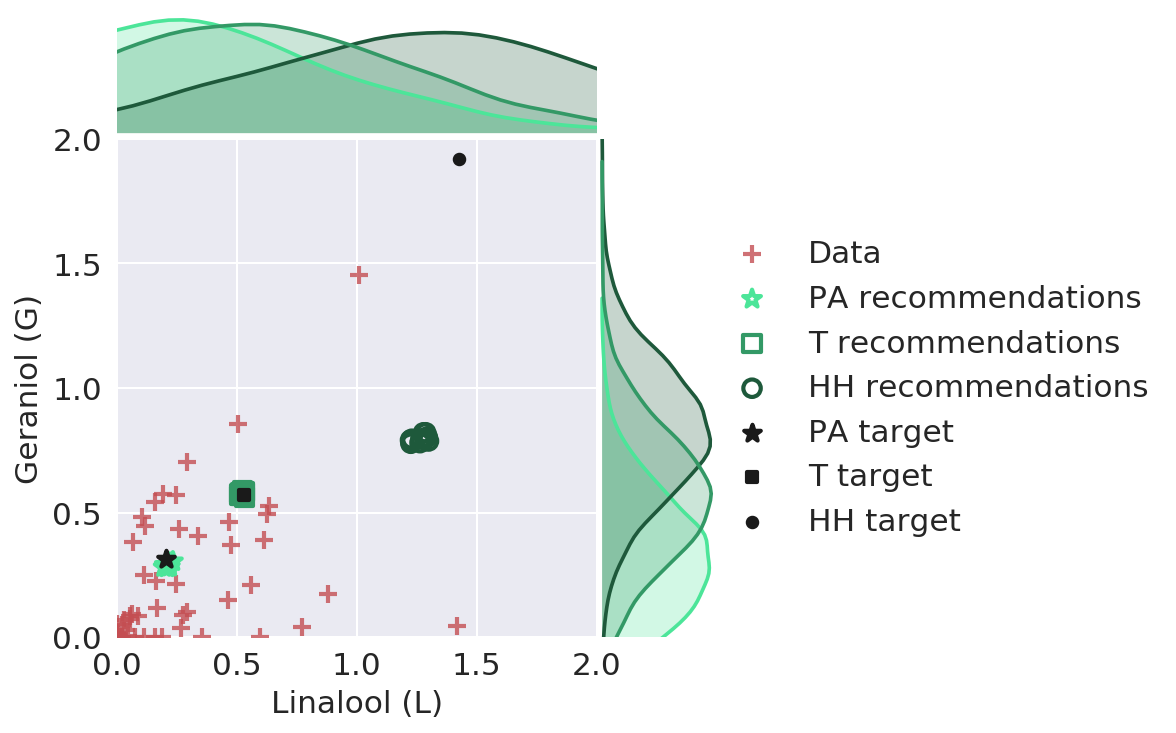}
	\end{center}
	\caption{\textbf{Linalool and geraniol predictions for \textsf{ART} recommendations for each of the beers (Fig.\ \ref{fig:example2}), showing full probability distributions (not just averages)}. These probability distributions (in different tones of green for each of the three beers) show very broad spreads, belying the illusion of accurate predictions and recommendations. These broad spreads indicate that the model has not converged yet and that many production levels are compatible with a given protein profile.}
	\label{FigHopless_recom_with_distr}
\end{figure}

\begin{figure}[!h]
	\begin{center}
		\includegraphics[width=1\columnwidth]{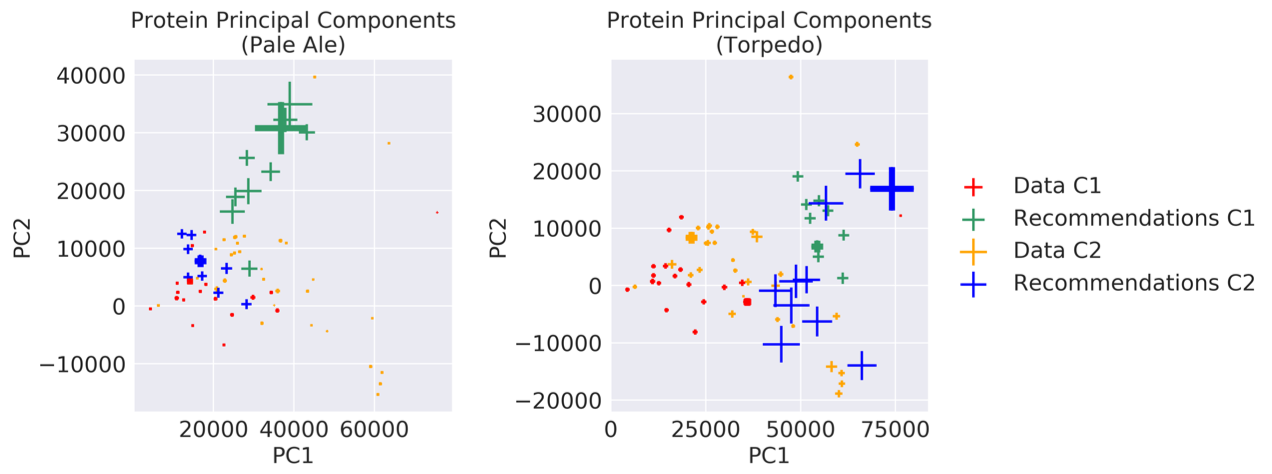}
	\end{center}
	\caption{\textbf{Principal Component Analysis (PCA) of proteomics data for the hopless beer project (Fig.~\ref{fig:example2})}, showing experimental results for cycle 1 and 2, as well as ART recommendations for both cycles. Cross size is inversely proportional to proximity to L and G targets (larger crosses are closer to target). The broad probability distributions spreads (Fig.~\ref{FigHopless_recom_with_distr}) suggest that recommendations will change significantly with new data. Indeed the protein profile recommendations for the Pale Ale changed markedly from DBTL cycle 1 to 2, even though the average metabolite predictions did not (Fig. 7, right column). For the Torpedo case, the final protein profile recommendations overlapped with the experimental protein profiles from cycle 2, although they did not cluster around the closest profile (largest orange cross), concentrating on a better solution according to the model. In any case, despite the limited predictive power afforded by the cycle 1 data, \textsf{ART} produces recommendations that guide the metabolic engineering effectively. For both of these cases, \textsf{ART} recommends exploring parts of the phase space such that the final protein profiles that were deemed close enough to the targets (in orange, see also bottom right of Fig.~\ref{fig:example2}) lie between the first cycle data (red) and these recommendations (green). In this way, finding the final target (expected around the orange cloud) becomes an interpolation problem, which is easier to solve than an extrapolation one.}
	\label{FigHoplessPCAs}
\end{figure}

\begin{table}[!h]
	\begin{tabular}{l|ll|ll}
		& \multicolumn{2}{c|}{Number of strains} & \multicolumn{2}{c}{Number of instances} \\ 
		& Cycle 1            & Cycle 2           & Cycle 1            & Cycle 2            \\ \hline
		Pathway 1 & 12                 & 11 (2)            & 50                 & 39 (6)               \\
		Pathway 2 & 9 (4)              & 10 (5)            & 31 (10)            & 30 (14)                \\
		Pathway 3 & 12                 & -                 & 35                 & -                 \\ \hline
		Total     & 33 (4)             & 21 (7)            & 116 (10)           & 69 (20) 
	\end{tabular}
	\caption{Total number of strains (pathway designs) and training instances available for the dodecanol production study\cite{Opgenorth2019}  (Figs.~\ref{FigDodecanol_P1}
		,~\ref{FigDodecanol_P2} and~\ref{FigDodecanol_P3}). Pathway 1, 2 and 3 refer to the top, medium and bottom pathways in Fig.~1B of \citet{Opgenorth2019}. Training instances are amplified by the use of fermentation replicates. Failed constructs (3 in each cycle, initial designs were for 36 and 24 strains in cycle 1 and 2) indicate nontarget, possibly toxic, effects related to the chosen designs. Numbers in parentheses () indicate cases for which no product (dodecanol) was detected.}
	\label{TabDodecanol_data}
\end{table}

\begin{figure}[!h]
	\begin{center}
		\includegraphics[width=1\columnwidth]{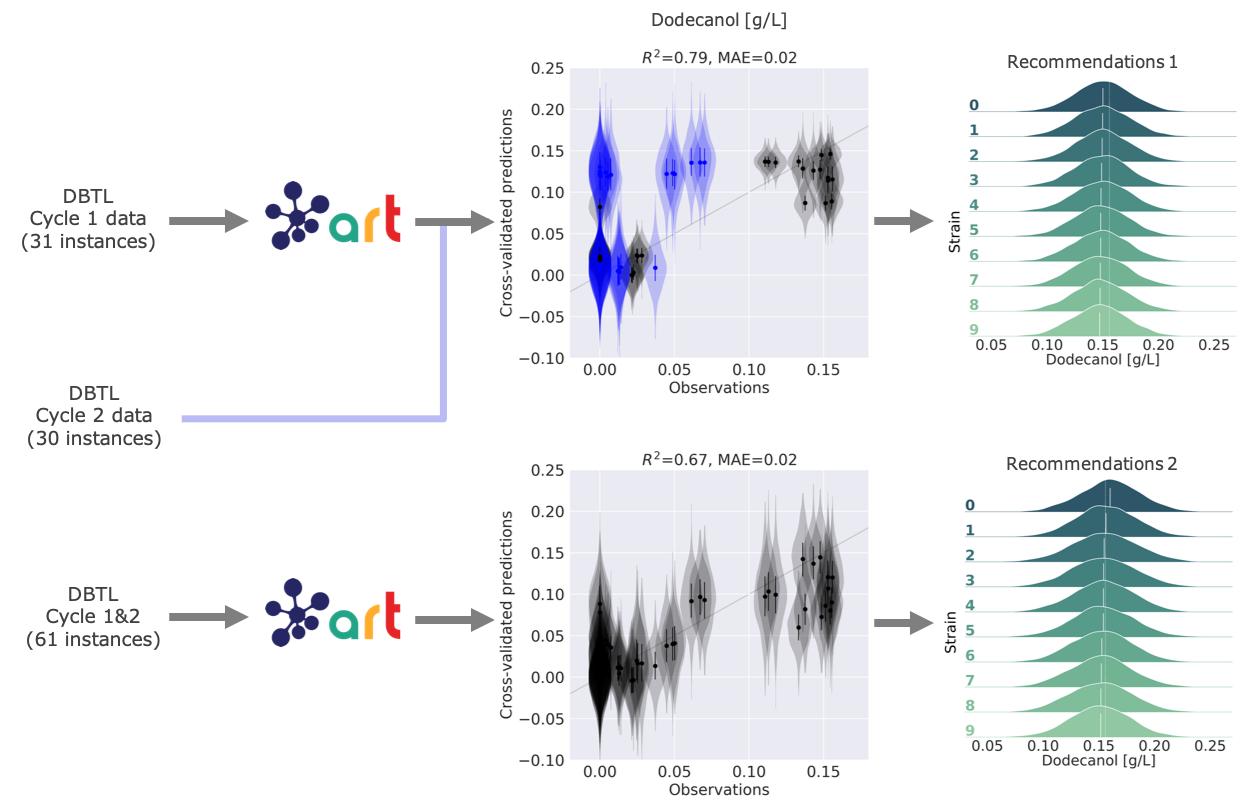}
	\end{center}
	\caption{\textbf{\textsf{ART}'s predictive power for the second pathway in the dodecanol production example is very limited.} Although cycle 1 data provide good cross-validated predictions, testing the model with 30 new instances from cycle 2 (in blue) shows limited predictive power and generalizability. As in the case of the first pathway (Fig.~\ref{FigDodecanol_P1}), combining data from cycles 1 and 2 improves predictions significantly.}
	\label{FigDodecanol_P2}
\end{figure}

\begin{figure}[!h]
	\begin{center}
		\includegraphics[width=1\columnwidth]{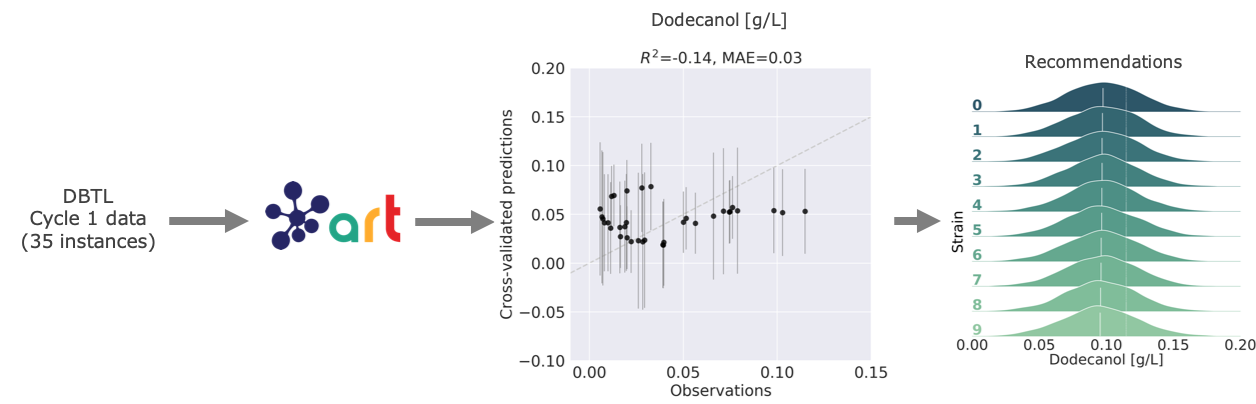}
	\end{center}
	\caption{\textbf{\textsf{ART}'s predictive power for the third pathway in the dodecanol production example is poor}. As in the case of the first pathway (Fig.~\ref{FigDodecanol_P1}), the predictive power using 35 instances is minimal. The low production for this pathway (Fig.\ 2 in \citet{Opgenorth2019}) preempted a second cycle.}
	\label{FigDodecanol_P3}
\end{figure}

\clearpage

\bibliography{ART}

\end{document}